\documentclass{emulateapj}
\usepackage{natbib}
\usepackage{apjfonts}

\newcommand{\msigma}{$M_{\rm BH}-\sigma_{\star}$}
\newcommand{\mlbulge}{$M_{\rm BH}-L_{\rm bulge}$}
\newcommand{\mlgalaxy}{$M_{\rm BH}-L_{\rm galaxy}$}
\newcommand{\mmbulge}{$M_{\rm BH}-M_{\rm bulge}$}
\newcommand{\mmstars}{$M_{\rm BH}-M_{\rm stars}$}
\newcommand{\mbh}{$M_{\rm BH}$}
\newcommand{\hst}{{\it HST}}
\newcommand{\sersic}{S\'{e}rsic}
\newcommand{\galfit}{{\sc Galfit}}

\shorttitle{Host Galaxies of RM AGNs}
\shortauthors{Bentz, et al.}

\received{}
\accepted{}

\begin{document}

\title{Black Hole - Galaxy Scaling Relationships for Active Galactic
  Nuclei with Reverberation Masses}

\author{ Misty~C.~Bentz\altaffilmark{1} and
Emily~Manne-Nicholas\altaffilmark{1}
}

\altaffiltext{1}{Department of Physics and Astronomy,
		 Georgia State University,
		 Atlanta, GA 30303, USA;
		 bentz@astro.gsu.edu}

\begin{abstract}

We have utilized high-resolution optical {\it Hubble Space Telescope}
images and deep, ground-based near-infrared images to examine the
host-galaxies of 37 active galactic nuclei with reverberation-based
black hole masses.  Using two-dimensional image decompositions, we
have separated the host galaxy from the bright central AGN, allowing a
re-examination of the \mlbulge\ and \mlgalaxy\ relationships and the
\mmbulge\ and \mmstars\ relationships using V-H color to constrain the
stellar mass-to-light ratio.  We find clear correlations for all of
these scaling relationships, and the best-fit correlations are
generally in good agreement with the sample of early-type galaxies
with \mbh\ from dynamical modeling and the sample of megamasers.  We
also find good agreement with the expectations from the Illustris
simulations, although the agreement with other simulations is less
clear because of the different black hole mass ranges that are
probed. \mlbulge\ is found to have the least scatter, and is therefore
the best predictor of \mbh\ among the relationships examined here.
Large photometric surveys that rely on automated analysis and forego
bulge-to-disk decompositions will achieve more accurate
\mbh\ predictions if they rely on \mmstars\ rather than \mlgalaxy.
Finally, we have examined $M_{\rm BH} / M_{\rm stars}$ and find a
clear trend with black hole mass but not galaxy mass.  This trend is
also exhibited by galaxies with \mbh\ from dynamical modeling and
megamaser galaxies, as well as simulated galaxies from Illustris,
rising from $~\sim 0.01$\% at $10^6$\,M$_{\odot}$ to $\sim 1.0$\% at
$10^{10}$\,M$_{\odot}$.  This scaling should be taken into account
when comparing galaxy samples that are not matched in \mbh.

\end{abstract}


\section{Introduction}

The discovery that nearly every massive galaxy hosts a supermassive
black hole in its nucleus is one of the lasting legacies of the Hubble
Space Telescope (\hst).  Direct measurements of the masses of these
black holes (\mbh), using luminous tracers inside the gravitational
potential of the invisible central massive object, have led to the
discovery of scaling relationships between the black holes and other
characteristics of their host galaxies.  This is true both for the
sample of mostly-quiescent galaxies with measurements of
\mbh\ from dynamical modeling of stars or gas (e.g.,
\citealt{kormendy13}) and for the sample of active galaxies that have
measurements of \mbh\ from reverberation mapping (e.g.,
\citealt{bentz15}).  

Direct black hole mass measurements are time and resource intensive,
and they are generally only applicable to galaxies that meet a
specific set of criteria.  For instance, reverberation mapping is only
applicable to broad-lined active galactic nuclei (AGNs), which are
rare in the local universe, whereas dynamical modeling is only
applicable when the black hole sphere of influence is resolved or
nearly so, which is generally only possible out to $\lesssim
100$\,Mpc.  The resource-intensive nature of these measurements as
well as the limitations on the applicability of each technique mean
that, in practical terms, the number of direct \mbh\ measurements that
may be accumulated over time is necessarily limited.  The scaling
relationships derived from these direct \mbh\ measurements, however,
provide valuable shortcuts for estimating \mbh\ based on less
resource-intensive measurements, such as the bulge stellar velocity
dispersion (the \msigma\ relationship;
\citealt{ferrarese00,gebhardt00}).  As such, direct \mbh\ measurements
and the scaling relationships that are based on them provide the
foundation for all other \mbh\ determinations, thereby providing
avenues to amass large samples for studying black hole and galaxy
co-evolution across galaxy types and at different look-back times
(e.g., \citealt{lapi14,heckman14,kelly12} and references therein).

Scaling relationships between the central black hole and the host
galaxy have also become important tools for critical testing of
cosmological simulations of dark matter halo mergers (e.g.,
\citealt{oogi16,degraf11,hopkins10}), numerical investigations of
candidate seed black holes (e.g.,
\citealt{shirakata16,volonteri09,lippai09}), cosmological modeling of
galaxy and black hole growth (e.g.,
\citealt{degraf15,kim11,bonoli09,miller06}), and investigations into
black hole feedback mechanisms (e.g.,
\citealt{steinborn15,kaviraj11,shabala11,ostriker10}).  Accurate
measurements of the host-galaxy characteristics of black holes with
direct \mbh\ measurements are therefore necessary and valuable.
Uncorrected biases or unmitigated scatter in the galaxy measurements can
adversely affect the accurate and precise calibration of widely-used
scaling relationships.

In this work, we focus on characterization of the host galaxies of
AGNs with reverberation-based \mbh\ measurements.  Using
high-resolution {\it HST} optical images and deep, ground-based
near-infrared images, we characterize the photometric properties of
the galaxies through two-dimensional image decompositions. We estimate
stellar masses based on photometric colors and widely-used $M/L$
prescriptions.  These results then allow us to recalibrate several
black hole-galaxy scaling relationships, and to investigate the black
hole mass to stellar mass fraction across the sample.

Throughout this work, we adopt a standard $\Lambda$CDM cosmology of
$H_0=72$\,km\,s$^{-1}$\,Mpc$^{-1}$ with $\Omega_{\Lambda}=0.7$ and
$\Omega_{M}=0.3$.

\section{Observations}

As part of our ongoing work with the reverberation sample of AGNs,
high-resolution medium-band $V$ observations were obtained with the
{\it Hubble Space Telescope} (\hst).  We have also recently collected
deep, ground-based near-infrared imaging for a number of these
galaxies at the WIYN observatory.  We restrict our analysis here to
the sample of 37 galaxies for which we have imaging in both the
optical and near infrared.  Table~\ref{tab:whirc.obs} lists the sample
and details of the observations, which we describe below.

\begin{figure*}
\plotone{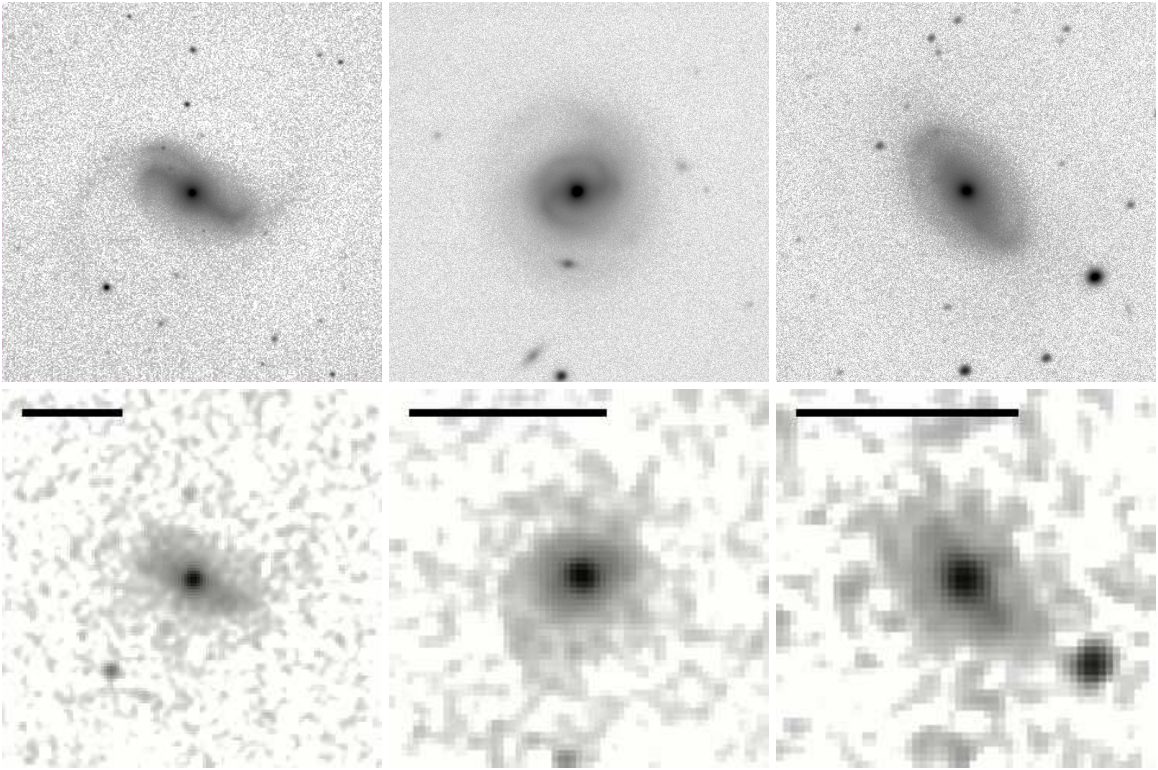}
\caption{ {\it Top row:} WIYN WHIRC $H$-band image negatives of the
  Seyfert galaxies Mrk\,79 ({\it left}), Mrk\,817 ({\it center}), and
  Zw\,229$-$015 ({\it right}).  All images are oriented with North up
  and East to the left. Note that we have successfully removed the
  fringing that is typical of this instrument in the $H$ band, and
  that there is a high level of morphological detail captured in each
  galaxy. {\it Bottom row:} 2MASS $H$-band image negatives of the same
  galaxies from the 2MASS Extended Source Catalog \citep{jarrett00a}.
  For each object, the displayed field of view for the 2MASS image
  matches the WIYN image, and the scale bars are 30\arcsec\ in length.
  The shallow depth and poorer spatial resolution of the 2MASS imaging
  loses many of the morphological details of the host galaxies, but
  these details are clearly captured in the WIYN images .}
\label{fig:2mass}
\end{figure*}

\subsection{Optical Imaging}

\hst\ imaging of the galaxies in our sample was acquired with the
following instrument configurations: the Advanced Camera for Surveys
(ACS) High Resolution Channel (HRC) through the F550M filter, the Wide
Field Planetary Camera 2 (WFPC2) with the F547M filter, and the Wide
Field Camera 3 (WFC3) through the F547M filter.  The medium-band $V$
filters were specifically chosen to avoid strong emission lines from
the AGN and to sample a flat portion of the underlying host-galaxy
spectrum.  The details of these observations and the post-processing
are described by \citet{bentz09b,bentz13}.

We also present here new WFC3 F547M images of eight galaxies in the
sample (HST GO-11661 and GO-13816, PI Bentz).  Three had not been
previously observed, while prior imaging of five galaxies with ACS HRC
provided a field of view ($29\farcs0 \times 25\farcs0$) that was too
narrow to capture their extended morphologies.  WFC3 provides a
$2\farcm7 \times 2\farcm7$ field of view that is well matched to the
galaxies in our sample, and a high spatial resolution with a pixel
scale of $0\farcs04$.  Each galaxy was observed for a single orbit,
with a 2-point dither pattern to fill in the gap between the
detectors.  At each point in the dither, a short and long exposure
were obtained.  The short exposures ensure an unsaturated measurement
of the bright central AGN at each position, while the long exposures
provide more depth for resolving the fainter, extended host galaxy.

The pipeline-reduced images were corrected for cosmic rays with
LACosmic \citep{vandokkum01}.  Taking advantage of the linear nature
of CCDs, we corrected for saturation of the AGN in the long exposures
by clipping out the saturated pixels in the long exposures and
replacing them with the same pixels from the short exposures taken at
the same dither position, scaled up by the exposure time ratio.  The
individual exposures were then drizzled to a common reference and
combined with AstroDrizzle.

\subsection{Near-Infrared Imaging}

Near-infrared imaging of 29 reverberation-mapped AGN host galaxies was
obtained between fall 2011 and spring 2013 with the WIYN
High-Resolution Infrared Camera (WHIRC) at the WIYN 3.5-m
telescope\footnote{The WIYN Observatory is a joint facility of the
  University of Wisconsin-Madison, Indiana University, the National
  Optical Astronomy Observatory and the University of Missouri.}.  The
camera is a Raytheon Virgo HgCdTe with a pixel scale of $0\farcs0986$
and a field of view of $202\arcsec \times 202\arcsec$.  While
broad-band $J$, $H$, and $Ks$ images were obtained for a subset of the
sample, the majority of the observations were devoted to $H$-band
images and we report those here.

The typical observing sequence involved many short observations of
each target with a generous dither pattern between observations.  This
allowed for the removal of strong fringing in the $H$ band, as well as
bad pixels and cosmic rays.

\begin{figure*}
\plotone{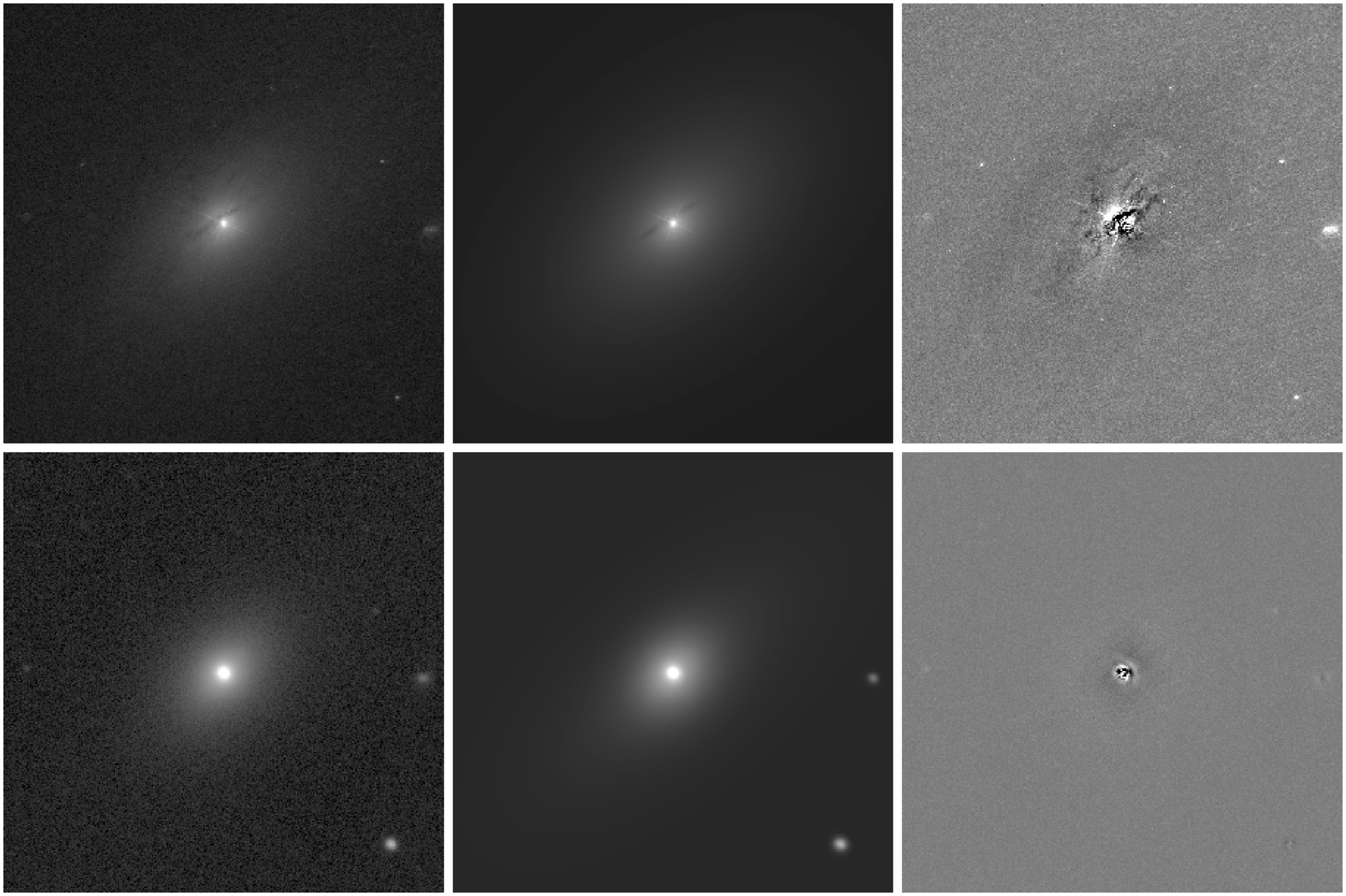}
\caption{Two-dimensional surface brightness decompositions of Mrk\,6,
  with the top row showing the medium-$V$ \hst\ image and the lower
  row showing the $H$-band WHIRC image. In the three panels of each
  row, we show the image ({\it left}), the model that was fit to the
  image ({\it center}), and the residuals after subtracting the model
  from the image ({\it right}).  The images and models are displayed
  with a logarithmic stretch, and the residuals are displayed with a
  linear stretch centered around zero counts.  The fields displayed
  are $1\arcmin \times 1\arcmin$ and are oriented with North up and
  East to the left. }
\label{fig:galfit}
\end{figure*}

Images were reduced in IRAF\footnote{IRAF is distributed by the
  National Optical Astronomy Observatory, which is operated by the
  Association of Universities for Research in Astronomy (AURA) under a
  cooperative agreement with the National Science Foundation.}
following standard procedures.  Strong fringing is a known problem for
$H-$ band images taken with WHIRC.  We were able to correct for this
effect by first median-combining a large number of dithered
observations of a target, with each image scaled by the median sky
level. Then we created a fringe mask from this combined image with the
IRAF task {\tt objmasks}.  Finally, we used the mask with the {\tt
  rmfringe} task to correct each image.  After correcting for
fringing, we computed the pixel offsets between dithered images,
subtracted the mean sky background, and shifted and combined all of
the images.  For the final image of each object, we added back the
average sky background that had been subtracted in the previous step,
to ensure that the image statistics would be properly handled in the
fitting process.  In Figure~\ref{fig:2mass} we show the final $H$-band
images for three of our targets in comparison to the Two Micron All
Sky Survey (2MASS;\citealt{skrutskie06}) $H$-band images for the same
galaxies.  The improvement in depth and spatial resolution provided by
the WHIRC images is immediately apparent, allowing for detection and
characterization of faint surface brightness features, as well as
better separation of distinct photometric components.

We supplemented this sample with \hst\ Near-Infrared Camera and
Multi-Object Spectrometer (NICMOS) observations of eight additional PG
quasars with the NIC2 camera through the F160W filter.  The details of
these observations are described by \citet{veilleux09}.  Drizzled and
combined images were downloaded from MAST.  For each image, we added
back the subtracted sky background as recorded in the header, and then
multiplied each image by the exposure time to return the image units
to counts.

\section{Surface Brightness Fits}

Two-dimensional surface brightness fits to the AGN host galaxy images
were carried out using the software \galfit\ \citep{peng02,peng10}.
\galfit\ allows the user to model surface brightness features with a
variety of analytical models.  We utilized the general
\citet{sersic68} profile to fit the various photometric components of each
galaxy.  This particular function has the form
\begin{equation}
\Sigma(r) = \Sigma_e \exp\left[-\kappa \left( \left(\frac{r}{r_e}\right)^{1/n} - 1\right)\right],
\end{equation}
where $\Sigma_e$ is the pixel surface brightness at the effective
radius $r_e$.  An exponential disk profile is simply a
\sersic\ profile with an index of $n=1$.  Bulges typically have $n>1$,
with the \citet{devaucouleurs48} profile being a special case with
$n=4$.  Bars, on the other hand, typically have $n<1$.  In the few
cases where a galaxy displayed a ring or a strong dust lane, we
utilized the truncation function to truncate the inner and/or outer
regions of a \sersic\ profile with $n=1$ to represent the ring.  For
these profiles, there are two quoted radii for each truncation
function, which are the break radius and the softening length.

Fits to many of the optical \hst\ images have already been published
by \citet{bentz09b,bentz13}.  Fits to the new WFC3 images were carried
out following similar procedures.  The point spread function (PSF) was
modeled by StarFit \citep{hamilton14} in an attempt to better account
for slight changes in the PSF width due to telescope
breathing. StarFit begins with a TinyTim PSF model \citep{krist93} and
attempts to match the telescope focus by fitting the PSF to a source
in the field.  Most of the galaxies did not have a suitable field star
in the frame to be used as a PSF model, so we used the StarFit model
derived from a star in the field of NGC\,3516 as the PSF model for all
eight galaxies.  While this provided a slight improvement over using
basic TinyTim PSF models, we still found that in several cases we
needed to supplement the PSF model with a narrow \sersic\ profile to
properly model the AGN in each galaxy nucleus.  Without the addition
of this component, the \sersic\ profiles for the bulge would run up to
an unrealistic index of $n>10$, and would often reach the default
maximum value of $n=20$.  Such profiles are extremely peaky with very
broad wings, effectively mimicking an unresolved point source and the
background sky.  Whenever this occurred, we added a narrow
(FWHM$\lesssim1$\,pixel) \sersic\ on top of the PSF model at the
location of the AGN.  The addition of this profile always resulted in
realistic values for all other model components in the image, allowing
all the model components to remain unfixed during the fitting. The
surface brightness profile of each galaxy was then fit with a bulge
and a disk model, with additional model parameters (such as a bar,
barlens, or ring) being added when necessary based on inspection of
the image and the residuals of the model. 

For the near-infrared WHIRC images, our fitting process began by
constructing a point spread function (PSF) image from an isolated
field star.  This first step involved analyzing a small portion of
each image centered on the star.  The background sky was modeled as a
tilted plane, and we fit multiple Gaussians to the star (typically
4-5) with unrestricted shape parameters and initial conditions of
widths graduated in size.  We also allowed for a single Fourier term
to provide an asymmetry in the light distribution of each Gaussian,
although we first arrived at a good set of model parameters before
turning this option on in the final fitting step.  The end result of
each PSF model image is a residual pattern (image minus model) that
does not retain any ``bulls-eye'' or other regular pattern, and is
simply consistent with noise.  These models, excluding the sky
component, were then used as the PSF images for fitting the galaxy.

The fits of the WHIRC $H$-band images were guided by the solutions
determined from the optical \hst\ images because of their superior
sensitivity and spatial resolution and lower sky background.  
The host-galaxy components in the WHIRC images were constrained to
have the same characteristic radii as had been found in the optical
images, scaled by the difference in the pixel scales.  Furthermore,
the indices of the \sersic\ profiles were constrained to the values
determined from the optical images.  In a few cases, unsatisfactory
fits of the WHIRC images led us to revisit and refine our
previously-published fits to an \hst\ image, and these new fits to the
optical images were then used to guide the fits to the WHIRC images.
In all cases, the final adopted fits in both the optical and
near-infrared bands agree, both in the number of photometric models
and their shape parameters.  The sky background was again fit
as a tilted plane, and the AGN and multiple field stars were fit with
the PSF image.  Field stars that were not fit were masked out.
Because of the non-photometric conditions throughout most of our WHIRC
observations, we adopted $H$-band magnitudes for as many field stars
as possible in each image from 2MASS.  The final zeropoint of each
WHIRC image was set by minimizing the differences between the reported
2MASS magnitudes and the \galfit\ magnitudes of the field stars.
Figure~\ref{fig:galfit} shows the \hst\ and WHIRC images for a typical
galaxy in our sample, Mrk\,6, as well as the surface brightness models
fit to each image and the residuals of the fits.

While the \sersic\ indices for the surface brightness components fit
to the optical \hst\ images were generally allowed to remain free
parameters, we followed a slightly different procedure for the
analysis of the eight PG quasars included in the sample of
\citet{veilleux09}.  At first, we intended to match the procedure
described by \citet{veilleux09} so that we could adopt the galaxy
magnitudes they report, but in the end we found that we preferred a
modified version of the procedure, and we thus re-fit all the NICMOS
images ourselves.

As with the other galaxies, we began with the optical images.  These
were all WFPC2 or ACS HRC images, and the small field of view did not
allow for a StarFit PSF model to be built from a suitably bright field
star.  The PSF was instead modeled with TinyTim \citep{krist93} and we
again found that we needed to add an additional narrow
\sersic\ component on top of the PSF to help avoid mismatch from
spacecraft breathing.  Each galaxy was fit with either a single
\sersic\ component with $n=4$, or with an exponential disk and a
\sersic\ component with $n=4$, depending on whether our previous fits
and those reported by \citet{veilleux09} found evidence for a disk or
not.  Faint field stars and galaxies in the images were fit
simultaneously with the AGN and its host galaxy, rather than masked
out.  Once a good fit was obtained in the optical image (all of which
have finer pixel scales and marginally larger fields of view than the
NIC2 camera), we turned to the fitting of the NICMOS images.  The
NICMOS PSF was modeled by TinyTim and was subsampled by a factor of 5,
and the galaxy model components were adopted from the optical images,
scaled to the proper size and held fixed during the fitting of the
NICMOS images, as we did with the WHIRC images.  The final parameters
for the adopted fits are listed in Table~\ref{tab:fits}.

\section{Galaxy Characteristics}

The \galfit\ models described in the previous section constrain the
observed magnitudes of the individual photometric components of the
host galaxies, which are generally, but not always, related to
kinematic components of the galaxy.  The fitting process that we
adopted allows for a direct comparison of the colors of individual
components (e.g., disks, bulges, bars), or the components can be
combined to investigate the total magnitude of a galaxy in each
passband as well as the overall galaxy color.

\subsection{Final Photometry}

The optical \hst\ magnitudes represent a few different medium-band $V$
filters rather than a true broad-band Johnson filter.  For each
object, we used {\sc synphot} and a reddened, redshifted
galaxy spectrum to determine the color difference between the filter
used for the optical \hst\ observations and a broad-band $V$ filter.
We adopted the elliptical galaxy template spectrum of \citet{kinney96}
for these calculations, but the use of an Sa or Sc galaxy spectrum
does not significantly change our results. The color differences are
small, $-0.05 < m_{V}-m_{HST} < 0.13$\, mag.  Similarly, we determined the
difference between the $F160W$ magnitudes from the NICMOS images and a
``true'' $H$-band magnitude using {\sc synphot} and the tabulated
passband for the $H$ filter provided on the WHIRC filters
webpage\footnote{https://www.noao.edu/kpno/manuals/whirc/filters.html}.
These corrections were slightly larger in magnitude and were always in
the same direction, showing a slight bias between the two filters,
$-0.15 < m_{H}-m_{F160W} < -0.12$\,mag.

The $V$ and $H$ magnitudes were then corrected for Galactic extinction
along the line of sight based on the values determined from the
\citet{schlafly11} recalibration of the \citet{schlegel98} dust map of
the Milky Way.  Table~\ref{tab:mags} gives the extinction-corrected
$V-$ and $H-$equivalent apparent magnitudes for the integrated
galaxies and for their bulges.

Based on our previous experience with \galfit\, as well as comparison
of our fitting results with those of \citet{veilleux09} for several of
the PG objects, we assume a typical uncertainty of $0.20$\,mag for the
integrated magnitudes of the galaxies.  We also assume a typical
uncertainty of $0.20$\,mag for the integrated magnitudes of the bulge
components on their own.  This is not to say that all the uncertainty
is in the bulges, but rather that \galfit\ does a good job of
recovering the total galaxy flux, even if there is some abiguity in
how the light is divided between the various photometric components.

\subsection{$V-H$ Colors}

The galaxies in our sample span a range of distances covering $0.0 < z
< 0.3$.  We used {\sc synphot} to determine $k$-corrections for each
galaxy in $V$ and $H$ so that we could compare $z=0$-equivalent
photometry and galaxy colors.  In $H$, the relatively flat spectral
energy distribution (SED) of a galaxy gives rise to small corrections
ranging from $-0.075 < k_H < -0.003$\,mag with a median of $k_H =
-0.028$\,mag.  In $V$, however, there is significantly more structure
to a galaxy SED, so the galaxies with the largest redshifts in our
sample ($z=0.1-0.3$) have $k_V > 0.2$\,mag.  The median is
significantly smaller, however, at $k_V = 0.061$\,mag.

After applying the $k$-corrections, we derive the $V-H$ colors, both
for the integrated light of the galaxy, which we report in
Table~\ref{tab:mass}, as well as for each individual component (e.g.,
bulge, bar, disk).  The range of integrated galaxy colors is $1.2 <
V-H < 3.3$, with a median color of $V-H=2.6$.  Most of the sample is
comprised of disk galaxies, and as expected, the $V-H$ colors of the
galaxy bulges are generally larger (redder) than the integrated colors
of the entire galaxies by a median value of 0.75\,mag.

\subsection{Distances}

To convert the observed magnitudes to luminosities, we estimated $D_L$
based on the apparent redshifts of the galaxies. We conservatively
adopt an uncertainty of 500\,km\,s$^{-1}$ in peculiar velocities for
each distance estimate, based on the distribution of peculiar
velocities derived by \citet{tully08}.  This works out to a 17\%
uncertainty at $z=0.01$, and decreases as $z$ increases.  We caution
that this uncertainty may still be significantly underestimated in the
case of individual galaxies.

Four of the nearest galaxies in the sample have distances in the
literature that were derived through other techniques.  We summarize
these distance measurements and their potential uncertainties in
\citet{bentz13}, but in brief, they were generally retrieved from the
Extragalactic Distance Database \citep{tully09} and derived from an
average of the distance moduli for galaxies within the same group.
The exception is NGC\,3227, where the distance measurement comes from
an analysis of the surface brightness fluctuations of NGC\,3226
\citep{tonry01}, with which it is interacting.


Adopted distances and their uncertainties are listed in
Table~\ref{tab:mags}.

\subsection{Stellar Masses}

We estimated the stellar mass-to-light ratio ($M/L$) of each galaxy
using the $V-H$ color and the relationships tabulated by
\citet{bell01} in their Table~1. Following their work, we assume solar
absolute magnitudes of $M_V = 4.82$ \citep{cox00} and $M_H = 3.37$
\citep{worthey94} and we derive the expected stellar mass in the $V$
and $H$ passbands, which are identical and are listed as ``$\log M_{\rm
  stars}$~(Bd01)'' in Table~\ref{tab:mass}.  The uncertainties on the
stellar masses are based on the propagated uncertainties in the
photometry and the distances.  The steeper dependency of $M/L$ on
bluer colors leads to uncertainties in the stellar masses that are
roughly twice as large when based on the luminosity in $V$ as for $H$.
To be conservative, we adopt the larger uncertainties based on $V$.

We also estimated the stellar mass-to-light ratio of each galaxy using
the relationships tabulated by \citet{into13}.  \citet{into13} used
updated population synthesis models and applied prescriptions to more
accurately account for thermally pulsing asymptotic giant branch
stars, which can strongly affect near-infrared photometry of galaxies.
We apply their ``dusty'' models (their Table~6) because we have not
attempted to correct for dust internal to the galaxies, and the
predominantly late types of the galaxies in our sample mean that they
cannot be considered ``dust free''.  Adopting the solar absolute
magnitudes of $M_V = 4.828$ and $M_H = 3.356$ derived by
\citet{into13} for consistency, we determined the expected stellar
masses in the $V$ and $H$ passbands.  Unlike the masses estimated from
the relationships of \citet{bell01}, the two passbands do not predict
exactly the same stellar masses, but the differences are only at the
1\% level.  We adopt the stellar masses based on the luminosity in
$V$, again because their larger propagated uncertainties make them a
more conservative choice, and we list these in Table~\ref{tab:mass} as
``$\log M_{\rm stars}$~(IP13)''.

The stellar masses predicted by the relationships of \citet{into13}
are typically a factor of 2.4 times smaller than those derived from
the relationships tabulated by \citet{bell01}, although for the
smallest $V-H$ colors in the sample, they can disagree by a factor of
$\sim4-5$.  Some of this difference can be attributed to the choice of
a ``diet'' Salpeter initial-mass-function (IMF) used by \citet{bell01}
and a Kroupa IMF adopted by \citet{into13}.  If we adjust the $M/L$
prescription of \citet{bell01} by $-0.15$\,dex to better match a
Kroupa IMF \citep{bell03}, then the agreement is better, although the
stellar masses predicted by the \citet{into13} $M/L$ relations are
still typically 1.7 times smaller than those predicted by
\citet{bell01}.  This difference agrees with the factor-of-two lighter
masses found by \citet{into13} when comparing their new stellar
population models with previous models, which they attribute to
updates like their treatment of thermally pulsing asymptotic giant
branch stars.  \citet{kormendy13} also find that the $M/L$ ratios
predicted by \citet{into13} are, on average, a factor of 1.34 smaller
than those predicted by the dynamics of the galaxy, although it is
unclear how much of the discrepancy may be attributed to dark matter.
Given the uncertainties in the methods, we report the results using
both the \citet{bell01} and the \citet{into13} prescriptions
throughout this work.

\begin{figure*}
\plotone{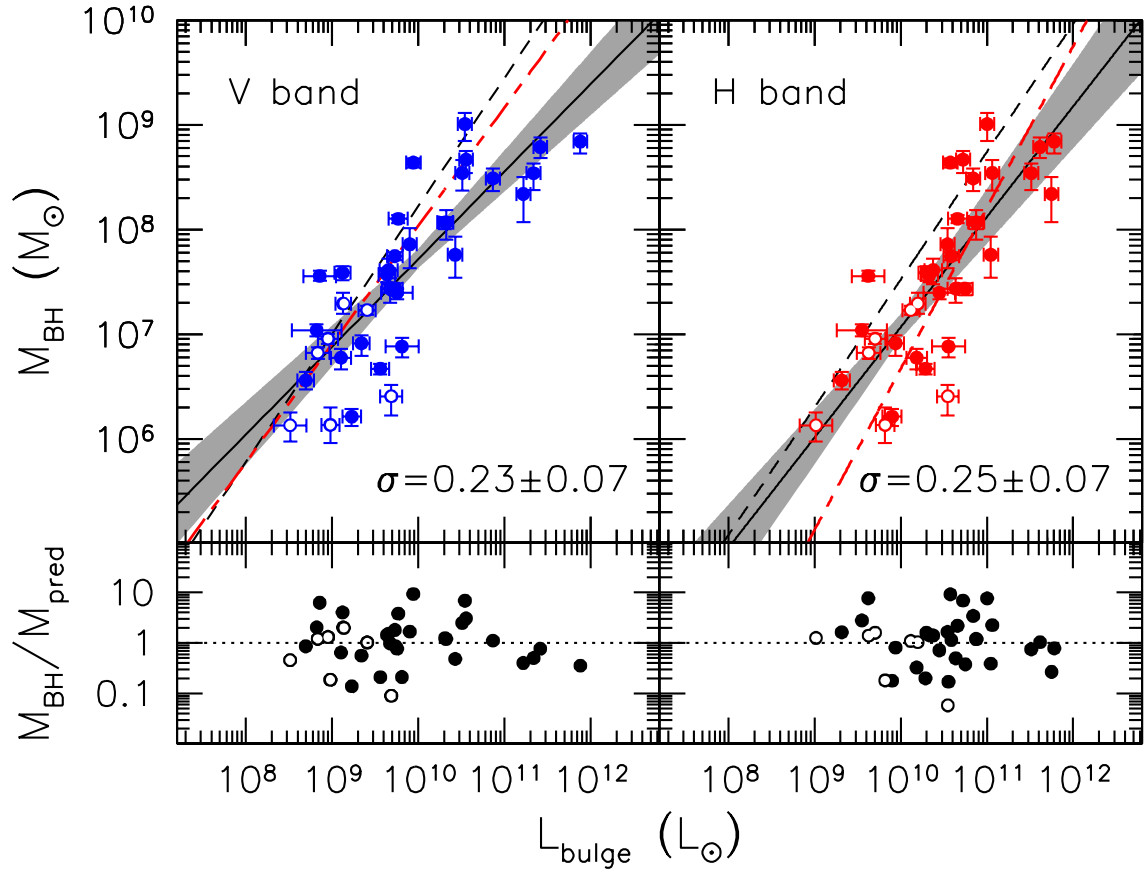}
\caption{Black hole mass as a function of the bulge luminosity of the
  galaxy, as determined from the $V$-band photometry ({\it left}) and
  the $H$-band photometry ({\it right}).  The solid lines show the
  best fits, while the gray shaded regions show the $1\sigma$
  uncertainties on the fits.  The scatter for the $V$-band
  relationship is formally smaller than that for the $H$-band
  relationship, but they are equivalent within the uncertainties.  The
  filled symbols denote the broad-line Seyfert 1s while the open
  symbols denote narrow-line Seyfert 1s, which appear to follow the
  same relationship and scatter. The black dashed lines show the best
  fits to the quiescent galaxy sample of \citet{kormendy13}, while the
  red long-short dashed lines show the best fits to the combined
  samples, including the $H$-band measurements of megamasers from
  \citet{lasker16}. The bottom panels show the distributions of
  measured \mbh\ relative to \mbh\ predicted by the best fit, as a
  function of $L_{\rm bulge}$. }
\label{fig:bulge}
\end{figure*}

\subsection{Black Hole Masses}

Black hole masses for all galaxies were drawn from the compilation of
reverberation-based masses in the AGN Black Hole Mass Database
\citep{bentz15}.  The basic technique of reverberation mapping
\citep{blandford82,peterson93} involves time-resolved
spectrophotometry collected over a long time baseline and with dense
time sampling (for nearby Seyferts, this typically amounts to daily
sampling over a baseline of a few months).  Variations in the continuum
flux are ``echoed'' in the broad emission lines, and the time delay
between the two is based on the light-travel time between the two
regions where the signals arise, namely the accretion disk and the
broad line region.

The black hole masses are determined as
\begin{equation}
M_{\rm BH} = f \frac{c \tau V^2}{G}
\end{equation}
\noindent where $c\tau$ is the measured time delay for a broad
emission line, $V$ is the velocity width of that same emission line,
and $G$ is the gravitational constant.  The factor $f$ is an
order-unity scaling factor that is necessary to account for the
generally unknown geometry and detailed kinematics of the broad
line region in the AGNs.  The value of $f$ ranges from
$2.8-5.5$ in the literature, with most current studies finding $f
\approx 4$.  We adopted the scaling factor of $\langle f \rangle =
4.3$ determined by \citet{grier13}.

\section{Discussion}

With the measurements of luminosities and masses derived in the
previous sections, we examine several black hole scaling relationships
here.  Linear regressions were carried out with a Bayesian approach
using the {\sc linmix\_err} algorithm \citep{kelly07}, which includes
measurement errors in both coordinates and a component of intrinsic,
random scatter.  The values and uncertainties that we report for
the slope, intercept, and scatter of each relationship are the median
values and $1\sigma$ widths of a large number of draws from the
posterior probability distribution for each term.

\subsection{Black Hole Mass -- Bulge Luminosity Relationship}

The relationship between black hole mass and bulge luminosity,
\mlbulge\, was one of the first black hole scaling relationships to be
discovered \citep{kormendy95}.  However, it was soon eclipsed by the
\msigma\ relationship \citep{ferrarese00,gebhardt00}, which was
initially reported to have a smaller intrinsic scatter and was
therefore viewed as being the more fundamental scaling relationship.
However, improvements in the black hole mass measurements, in
particular, have led to much tighter \mlbulge\ relationships in recent
years where the reported scatter is similar to that of the
\msigma\ relationship \citep{marconi03,gultekin09}.  These studies
have tended to focus on bulge-dominated galaxies while neglecting the
late-type galaxies common among local Seyfert hosts.

A notable exception, however, is \citet{wandel02}, who drew photometry
from the literature to investigate the \mlbulge\ relationship for AGN
host galaxies with black hole masses from reverberation mapping.  A
homogeneous reanalysis of the AGN black hole masses by
\citet{peterson04} combined with consistent bulge photometry derived
from high quality \hst\ imaging and galaxy photometric decompositions
allowed \citet{bentz09a} to update the results of \citet{wandel02},
finding that \mlbulge\ for disk-dominated active galaxies is similar
in form and scatter to that of bulge-dominated galaxies with
predominantly quiescent black holes and masses derived from dynamical
modeling.

Here, we are able to improve upon the results of \citet{bentz09a} by
extending the sample to lower black hole masses, increasing the number
of galaxies included in the fit by 40\%, and by examining the
relationship in both the optical and the near-infrared.  This last
point is an important addition because it allows for the effects of
dust and recent star formation on the photometry to be mitigated.

For each galaxy, we identified the photometric component most
consistent with the expected properties of a bulge.  In particular, we
looked for a round ($0.7 \lesssim q \lesssim 1.0$) photometric
component with \sersic\ index $n>1.0$ and $r < r_{\rm disk}$.  In one
instance (Mrk\,509), there was no such model component and so we do
not include it here in the analysis of galaxy bulges.  Mrk\,509 is
thus consistent with either a bulgeless disk galaxy or a disk galaxy
with a compact bulge that we could not separate from the central AGN.
Some of the PG quasars, on the other hand, were modeled by a single
spheroidal component which we include as a ``bulge'' here.  We do not
attempt to discriminate between pseudobulges and classical bulges
because we have limited kinematic information regarding the bulges of
these galaxies.  Numerous studies have shown that pseudobulge
identification can be extremely uncertain when it is based solely on
photometric information (e.g., \citealt{lasker14a,kormendy04}).  In
the $V$ band, we find the best-fit relationship between the black hole
mass and bulge luminosity to be:
\begin{equation}
\log \frac{M_{\rm BH}}{M_{\odot}} = (0.84 \pm 0.10) \log \left(\frac{L_{\rm V,bulge}}{10^{10}L_{\odot}}\right) + (7.71 \pm 0.08)
\end{equation}
\noindent with a typical scatter of $(0.23 \pm 0.06)$\,dex.  This is
similar to the slope found by \citet{bentz09a} using a smaller number
of galaxies in the reverberation sample and covering a smaller range
of black hole masses.  The scatter is much decreased, however, from
$\sim 0.4$\,dex to $0.23$\,dex.

In the $H$ band, we find a best-fit relationship of:
\begin{equation}
\log \frac{M_{\rm BH}}{M_{\odot}} = (1.05 \pm 0.14) \log \left(\frac{L_{\rm H,bulge}}{10^{10}L_{\odot}}\right) + (7.06 \pm 0.11)
\end{equation}
\noindent with a typical scatter of $(0.25 \pm 0.07)$\,dex.
Surprisingly, the scatter in the near-infrared relationship is
statistically equivalent to that of the optical relationship,
suggesting that dust and/or recent star formation are not strong
contributors to the intrinsic scatter in the relationship.  As
previously mentioned, however, there is still room for improvement in
the distances, so it is likely that the scatter in both the optical
and near-infrared relationships could be further decreased in the
future through efforts to determine distances that do not rely on the
galaxy redshift.

We display these relationships in Figure~\ref{fig:bulge}.  The solid
line shows the best fit, while the gray shaded regions show the
uncertainties in the fit.  We denote broad-line Seyferts 1s (BLS1s)
with filled circles and narrow-line Seyferts 1s (NLS1s) with open
circles.  We follow the original definition of \citet{osterbrock85}
and select NLS1s in cases where the broad H$\beta$ emission line has
FWHM$<2000$\,km\,s$^{-1}$.  While the NLS1s tend to be associated with
lower-mass black holes in lower-luminosity bulges, they exhibit the
same scatter and general scaling relationship as the BLS1s.  Some
studies of NLS1s with black hole estimates have shown them to be
significantly undermassive relative to BLS1s (e.g.,
\citealt{mathur12}), but we see no strong tendency for NLS1s to be
undermassive relative to the other reverberation-mapped AGNs included
here.

\begin{figure*}
\plotone{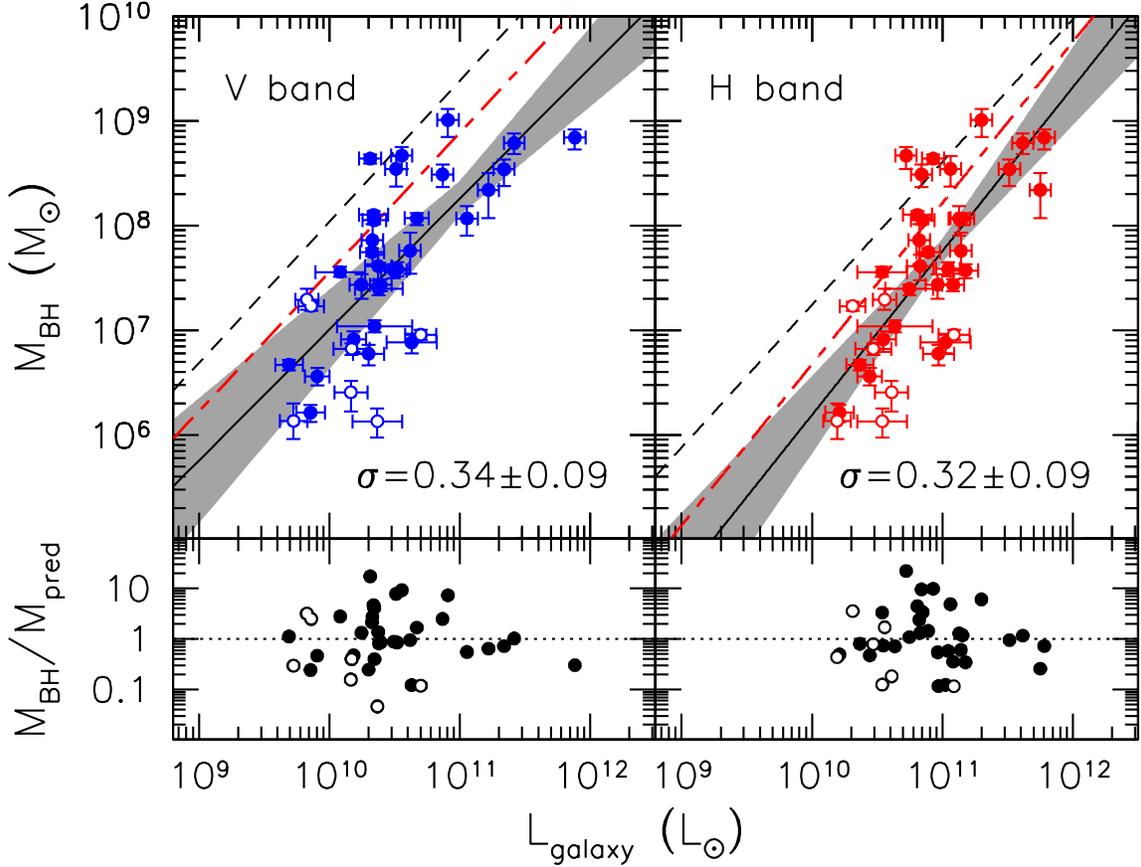}
\caption{Black hole mass as a function of the total luminosity of the
  galaxy, as determined from the $V$-band photometry ({\it left}) and
  the $H$-band photometry ({\it right}).  The solid lines show the
  best fits, while the gray shaded regions show the $1\sigma$
  uncertainties on the fits.  The scatter for the $H$-band
  relationship is formally smaller than that for the $V$-band
  relationship, but they are equivalent within the uncertainties. The
  black dashed lines show the best fits to the quiescent galaxy sample
  of \citet{kormendy13}, while the red long-short dashed lines show
  the best fits to the combined samples, including the $H$-band
  measurements of megamasers from \citet{lasker16}.  The bottom panels
  show the distributions of measured \mbh\ relative to \mbh\ predicted
  by the best fit, as a function of $L_{\rm galaxy}$.}
\label{fig:galaxy}
\end{figure*}

\citet{kormendy13} report a near-infrared \mlbulge\ relationship in
the 2MASS $K_S$ band for quiescent galaxies that are ellipticals or
contain classical bulges, and for which black hole masses have been
determined through dynamical modeling.  They find a slightly steeper
slope of $1.21$ and a scatter of $0.31$\,dex, both of which are
consistent within the errors with our finding for the active galaxy
sample in $H$. The slightly higher intercept for their sample compared
to ours is increased by the color difference between the $H$ and $K$
bands, given that galaxies are typically somewhat brighter in $K$ than
$H$.

While \citet{kormendy13} do not report a fit to the
\mlbulge\ relationship in $V$, they do tabulate bulge absolute
magnitudes in $V$.  We fit the $V$-band relationship matching their
accepted sample and adopted uncertainties and find a slope that agrees
with their value reported for the $K_s$ band, which is steeper than
the slope that we find in $V$ for the active galaxy sample.  The
intercept is also somewhat higher, although the fit to their sample
agrees with our findings for the active galaxy sample at the low-mass
end.  The fits to the \citet{kormendy13} sample are shown as black
dashed lines in Figure~\ref{fig:bulge}. It is important to note that
the active and quiescent samples primarily probe different regions of
parameter space in this plot: the active galaxy sample is heavily
dominated by galaxies with $M_{\rm BH} < 10^8 M_{\odot}$, while the
vast majority of galaxies in the quiescent sample have $M_{\rm BH} >
10^8 M_{\odot}$.

\citet{lasker16} report deep $H$-band imaging and surface brightness
decompositions for a sample of 9 megamaser galaxies with accurate
black hole masses.  We find that the megamasers are contained wholly
within the scatter of the active galaxy sample presented here in the
$H$ band.  With the good agreement between the active galaxies, the
megamasers, and the quiescent galaxy sample, we therefore refit the
\mlbulge\ relationship in $H$ with all three samples combined.  Based
on the typical galaxy properties in 2MASS reported by
\citet{jarrett00b}, we adopt $\langle H-K_s \rangle = 0.3$\,mag for
the quiescent sample, which should account for any average color
offset between the two filters (although we note that the scatter in
$H-K_s$ values is typically $\sim 0.2$\,mag, even for galaxies with a
specific morphological type). The best fit is:
\begin{equation}
\log \frac{M_{\rm BH}}{M_{\odot}} = (1.31 \pm 0.09) \log \left(\frac{L_{\rm H,bulge}}{10^{10}L_{\odot}}\right) + (7.27 \pm 0.08)
\end{equation}  
\noindent with a typical scatter of $0.26\pm0.05$\,dex.  While
\citet{lasker16} do not report $V$-band measurements for the megamaser
sample, we can investigate the \mlbulge\ relationship in $V$ for the
active and quiescent samples combined.  When we do, we find a best fit
of:
\begin{equation}
\log \frac{M_{\rm BH}}{M_{\odot}} = (1.13 \pm 0.08) \log \left(\frac{L_{\rm V,bulge}}{10^{10}L_{\odot}}\right) + (8.04 \pm 0.06)
\end{equation}
\noindent with a typical scatter of $(0.24 \pm 0.05)$\,dex.  These
fits are displayed as the red long-dashed lines in
Figure~\ref{fig:bulge}. In both the $H$ and $V$ bands, the best fit
for the combined sample has an almost identical scatter to that found
for the active sample alone, even though the combination of the
samples more than doubles the number of points being fit and extends
the range of \mbh\ by an order of magnitude.  This may indicate that
the galaxies in all three samples are drawn from the same parent
population.

\subsection{Black Hole Mass -- Galaxy Luminosity Relationship}

We also examined the relationship between black hole mass and total
luminosity of the host galaxy.  We find a clear correlation between
these two measurements, in both the optical and the near-infrared.
The best-fit relationships are found to be:
\begin{equation}
\log \frac{M_{\rm BH}}{M_{\odot}} = (1.25 \pm 0.22) \log \left(\frac{L_{\rm V,galaxy}}{10^{11}L_{\odot}}\right) + (8.26 \pm 0.16)
\end{equation}
\noindent in the $V$ band, with a typical scatter of $(0.34\pm0.09)$\,dex, and
\begin{equation}
\log \frac{M_{\rm BH}}{M_{\odot}} = (1.56 \pm 0.24) \log \left(\frac{L_{\rm H,galaxy}}{10^{11}L_{\odot}}\right) + (7.75 \pm 0.10)
\end{equation}
\noindent in the $H$ band, with a typical scatter of $(0.32\pm0.09)$\,dex.  

While the scatter is somewhat higher than that of the
\mlbulge\ relationship, the fact that there is still a relatively
tight relationship found when the total galaxy luminosity is used (see
Figure~\ref{fig:galaxy}) suggests that bulge/disk decompositions can
be avoided when estimating black hole masses from broad-band
photometry of disk galaxies, but with a loss of some accuracy.
This may be of particular interest for large photometric surveys that
are operational or coming online soon (e.g., LSST), where automated
measurements will be key to making sense of the large datasets that
will be produced.

Our best-fit relationships for active galaxies may again be compared
to the \citet{kormendy13} sample of quiescent galaxies.  The best-fit
relationships based on their tabulated measurements in $V$ and $K_s$
have similar slopes and scatter to our findings, but their intercepts
are significantly higher.  This appears to stem from the differences
in morphology among their sample and ours, as well as the different
ranges of \mbh\ between the two samples.  While the intercepts for the
\mlbulge\ relationships traced by the active galaxy sample show good
agreement with the quiescent galaxies, 2/3 of the galaxies in the
\citet{kormendy13} sample are ellipticals.  Thus, the
\mlgalaxy\ relationships for their sample are very similar to the
\mlbulge\ relationships, because 2/3 of the points between them are
exactly the same.
On the other hand, the active galaxy sample is dominated by later-type
galaxies where the bulge contributes a smaller fraction of the
integrated galaxy light, and so the best-fit \mlbulge\ and
\mlgalaxy\ relationships that we find for the active galaxies are
quite different from each other.

We looked at the bulge-to-total ratios for the active galaxy sample
and investigated whether splitting the sample into ``early'' (B/T >
0.5) and ``late'' (B/T < 0.5) types uncovered any offsets or
separations among the sample that may lead to better agreement with
the quiescent galaxy sample.  The only obvious difference between
these two subsamples is that the ``early'' types have more massive
black holes than the ``late'' types, and so a cut in B/T is similar to
a cut in \mbh\ and does not improve the agreement.  As before, we also
investigated the location of the \citet{lasker16} megamasers and find
that they are wholly contained within the $H$-band scatter of the
active galaxy sample.  If we again combine the active, quiescent, and
megamaser samples as before, we find best fits of:
\begin{equation}
\log \frac{M_{\rm BH}}{M_{\odot}} = (1.33 \pm 0.17) \log \left(\frac{L_{\rm V,galaxy}}{10^{11}L_{\odot}}\right) + (8.89 \pm 0.13)
\end{equation}
\noindent in the $V$ band, with a typical scatter of $(0.55\pm0.09)$\,dex, and
\begin{equation}
\log \frac{M_{\rm BH}}{M_{\odot}} = (1.54 \pm 0.18) \log \left(\frac{L_{\rm H,galaxy}}{10^{11}L_{\odot}}\right) + (8.22 \pm 0.08)
\end{equation}
\noindent in the $H$ band, with a typical scatter of
$(0.52\pm0.08)$\,dex.  Thus, while the scatter is significantly
increased when galaxies of all morphological types are treated
equally, it is likely more representative of the true uncertainty on
black hole mass estimates from the total galaxy luminosity.  

\begin{figure*}
\plotone{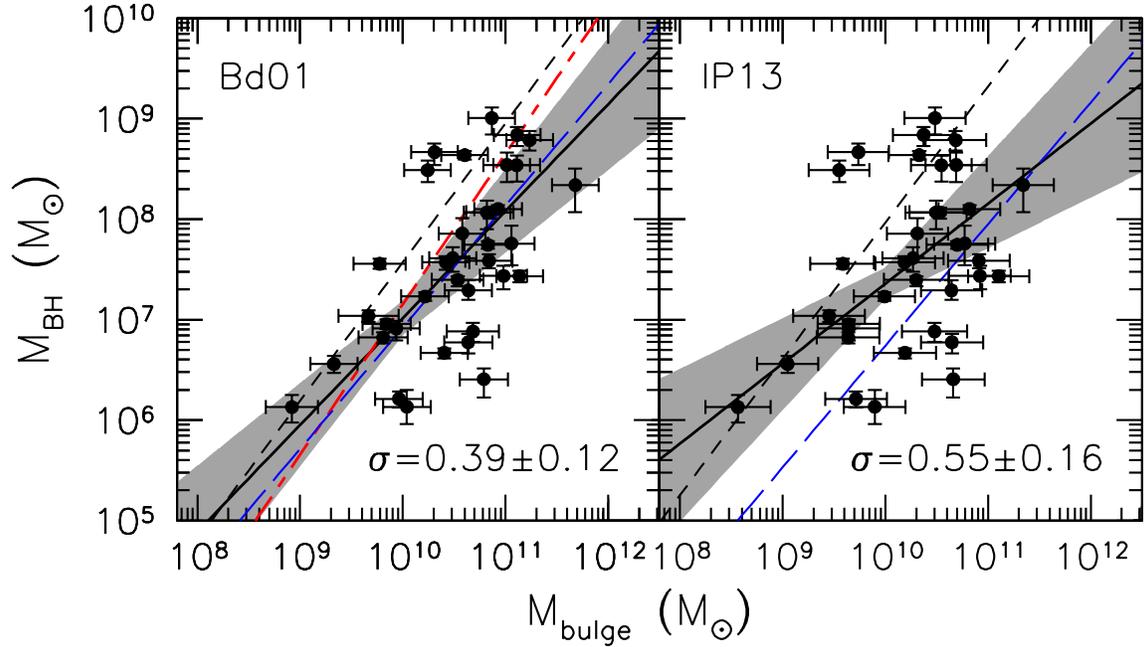}
\caption{Black hole mass as a function of the bulge stellar mass,
  where stellar mass is calculated based on the $V-H$ color and the
  $M/L$ prescriptions of \citet{bell01} ({\it left}) and
  \citet{into13} ({\it right}).  The solid lines and gray regions show
  the best-fit lines and $1\sigma$ uncertainties on the fits.  The
  dashed lines show the best fit for the sample of quiescent galaxies
  tabulated by \citet{kormendy13}.  The blue long-dashed lines show
  the best fit determined by \citet{sijacki15} for galaxies from the
  Illustris simulation. The red long-short dashed line is the best fit
  when the active galaxies, quiescent galaxies, and megamaser samples
  are combined.}
\label{fig:bstmass}
\end{figure*}

\subsection{Black Hole Mass -- Bulge Stellar Mass Relationship}

The relationship between black hole mass and bulge stellar mass is
expected to be the physical basis for the \mlbulge\ relationship,
where bulge light traces mass.  A variety of methods have been used to
investigate this relationship in the past, often with the aim of
decoupling the \mmbulge\ relationship from any dependence on the
\mlbulge\ relationship so they can be studied independently.

For example, \citet{magorrian98} carried out axisymmetric dynamical
models to constrain the bulge mass and the black hole mass
simultaneously.  \citet{marconi03} measured effective bulge radii from
2MASS imaging for quiescent galaxies with dynamical black hole masses.
The bulge radii were combined with $\sigma_*$ to predict $M_{\rm
  bulge}$ under the assumption that bulges behave similarly to
isothermal spheres.  \citet{haring04}, on the other hand, numerically
solved the spherical Jeans equation while matching published
luminosity and $\sigma_*$ profiles for quiescent galaxies with
dynamical black hole masses.

We can examine this relationship for active galaxies by estimating the
bulge stellar mass from its optical$-$near-infrared color and the
$M/L$ prescriptions described above.  The best-fit relationship
between the black hole mass and the stellar mass of the bulge, based
on the \citet{bell01} $M/L$ predictions, is found to be:
\begin{equation}
\log \frac{M_{\rm BH}}{M_{\odot}} = (1.06\pm 0.24) \log \left(\frac{M_{\rm bulge}}{10^{10}M_{\odot}}\right) + (7.02 \pm 0.17)
\end{equation}
\noindent with a typical scatter of $(0.39 \pm 0.12)$\,dex.

If we estimate $M/L$ using the prescriptions of \citet{into13}, we
find the best fit to be:
\begin{equation}
\log \frac{M_{\rm BH}}{M_{\odot}} = (0.80 \pm 0.30) \log \left(\frac{M_{\rm bulge}}{10^{10}M_{\odot}}\right) + (7.36 \pm 0.15)
\end{equation}
\noindent with a typical scatter of $(0.55 \pm 0.16)$\,dex.  These
relationships are displayed in Figure~\ref{fig:bstmass}.

\begin{figure*}
\plotone{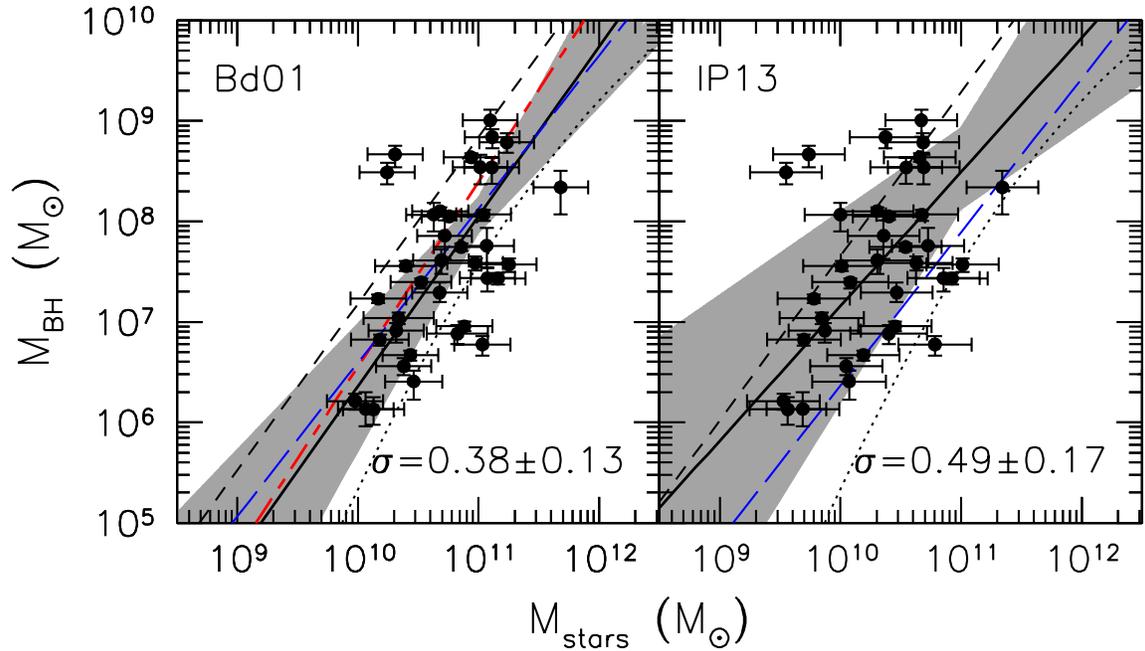}
\caption{Black hole mass as a function of the total stellar mass of
  the galaxy, where stellar mass is calculated based on the $V-H$
  color and the $M/L$ prescriptions of \citet{bell01} ({\it left}) and
  \citet{into13} ({\it right}).  The solid lines and gray regions show
  the best-fit lines and $1\sigma$ uncertainties on the fits.  The
  \citet{bell01} $M/L$ prescriptions lead to a tighter relationship
  with a somewhat steeper slope.  The dashed lines show the best fit
  for the sample of quiescent galaxies tabulated by
  \citet{kormendy13}.  The dotted line is the predicted ``unbiased''
  relationship of \citet{shankar16}, which agrees extremely well with
  our measurements if we adopt $M/L$ from \citet{bell01} and $f=1$
  rather than $f=4.3$ for $M_{\rm BH}$ (as \citealt{shankar16}
  recommend). The blue long-dashed lines show the relationship found for
  disk galaxies in the Illustris simulation \citep{mutlupakdil18},
  which agrees extremely well with our measurements of \mbh\ and
  $M_{\rm stars}$ when the $M/L$ prescriptions of \citet{bell01} are
  adopted.  The red long-short dashed line is the best fit when the
  active galaxies, quiescent galaxies, and megamaser samples are
  combined.}
\label{fig:stmass}
\end{figure*}

For a direct comparison with the quiescent galaxy sample, we
recalculated the bulge masses based on the absolute $V$ magnitudes of
the bulges and the $V-K_s$ colors tabulated by \citet{kormendy13} with
the $M/L$ prescriptions of both \citet{bell01} and \citet{into13}.
Because \citet{kormendy13} only provide an integrated $V-K_s$ color
for each galaxy, we note that we would expect there to be a bias in
the bulge masses derived for the disk galaxies in their sample because
of the different colors of bulges and disks.  The best-fit
relationships for the quiescent galaxies are shown as the black dashed
lines in Figure~\ref{fig:bstmass}.

While the active galaxy sample displays a linear relationship between
\mbh\ and bulge stellar mass, the quiescent galaxy relationships are
quite a bit steeper.  The two samples agree better using the $M/L$
prescriptions of \citet{bell01}, although both prescriptions show
agreement between the samples at the low mass end.

The megamaser sample of \citet{lasker16} reports $M_{\rm bulge}$ based
on near-infrared {\it HST} and ground-based imaging and the $M/L$
prescriptions of \citet{bell03}, which allows for a simple comparison
with our results.  We again find that all 9 megamasers are contained
wholly within the scatter of the active galaxy sample, with no
apparent offsets in bulge mass or black hole mass.

Noting that there is good agreement between the active, quiescent, and
megamaser samples, we also fit the \mmbulge\ relationship with all
three samples combined.  Assuming the \citet{bell01} $M/L$
prescriptions, the best-fit relationship is:
\begin{equation}
\log \frac{M_{\rm BH}}{M_{\odot}} = (1.50 \pm 0.13) \log \left(\frac{M_{\rm bulge}}{10^{10}M_{\odot}}\right) + (7.16 \pm 0.11)
\end{equation}
\noindent with a scatter of ($0.27 \pm 0.06$)\,dex.  The red
long-short dashed line in the left panel of Figure~\ref{fig:bstmass}
displays this fit.  While the slope is quite a bit steeper than that
found for the active galaxy sample, they only disagree at the $\sim
1\sigma$ level.  Furthermore, there is good agreement with the gray
shaded region (which denotes the uncertainty on the fit to the active
sample alone) over the range sampled by the active galaxies.  This
appears to indicate that all three samples may be drawn from the same
parent population of galaxies.

We also compared our results to those of simulated galaxies.  Some
caution must be taken when interpreting such comparisons, because
cosmological galaxy simulations are generally tuned to match a set of
observables.  For example, the slope of the \mmbulge\ relationship is
not expected to be affected by such tuning, but the intercept is.
Furthermore, there is no agreement on the best way to separate the
bulges of late-type galaxies from their disks in simulations, where
the resolution is often a limiting factor, so the simulated galaxies
are either compared to samples of massive early-type galaxies where
$M_{\rm bulge} \approx M_{\rm galaxy}$ (e.g., \citealt{steinborn15,schaye15})
or a prescription is applied to estimate the bulge contribution.  

\citet{sijacki15} used the high-resolution hydrodynamical Illustris
simulations to explore the predicted \mmbulge\ relationship for 
galaxies.  The total stellar mass within the stellar half-mass radius
was used as a proxy for the bulge mass.  This simplification does not
take into account different bulge mass fractions of galaxies, nor the
fact that some galaxies may not have a bulge at all.  Additionally,
the Illustris simulations assumed a Chabrier IMF, which can be
compared to a ``diet'' Salpeter like that employed by \citet{bell01}
by adding 0.093\,dex \citep{gallazzi08}.  To compare with a Kroupa IMF
like that employed by \citet{into13}, on the other hand, we subtracted
0.057\,dex \citep{bell03,herrmann16}.  The best-fit relationship of
\citet{sijacki15}, with the IMF scaled appropriately, is displayed as
the blue long-dashed lines in Figure~\ref{fig:bstmass}.  With a reported
slope of $1.21$, it is in good agreement with our findings, especially
when we adopt the \citet{bell01} $M/L$ prescriptions.

The large-volume Horizon-AGN simulations, which adopt a Salpeter IMF,
were analyzed by \citet{volonteri16}.  To separate the bulge
contribution, they tried various prescriptions, including examining
the kinematics and also adopting a double \sersic\ model for each
galaxy, where the indices for the two \sersic\ profiles were chosen to
be [1.0, 1.0], [1.0, 4.0], or [1.0, 1.0 or 4.0]. The slope of the
relationship based on these various prescriptions ranges from
$0.75-1.05$, which is in good agreement with our findings for the
active galaxies using either the \citet{bell01} or \citet{into13}
$M/L$ prescriptions, although it is somewhat in tension with our
results for the combined active, quiescent, and megamaser samples.
This tension may result from incompleteness in the Horizon-AGN
simulation for black holes with $M_{\rm BH} \lesssim 2\times
10^7$\,M$_{\odot}$, which is the region probed by many of the active
galaxies.

\begin{figure*}
\plotone{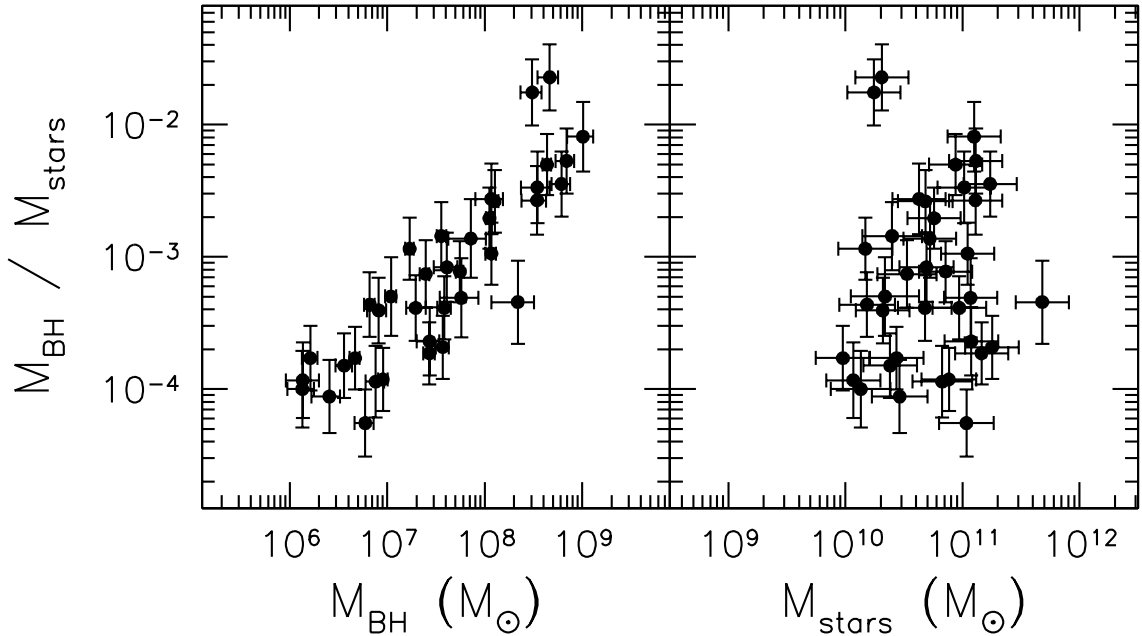}
\caption{Black hole mass fraction as a function of black hole mass
  ({\it left}) and as a function of galaxy stellar mass ({\it right}).
  There is no correlation seen between black hole mass fraction and
  galaxy stellar mass, but there appears to be a strong correlation
  between black hole mass fraction and black hole mass, with more
  massive black holes commanding a larger mass fraction.}
\label{fig:mbhfrac}
\end{figure*}

\subsection{Black Hole Mass -- Galaxy Stellar Mass Relationship}

In the same way, we can examine the best-fit relationship between the
black hole mass and the total stellar mass of the galaxy.  When we
adopt the \citet{bell01} $M/L$ predictions, we find a best fit of:
\begin{equation}
\log \frac{M_{\rm BH}}{M_{\odot}} = (1.69 \pm 0.46) \log \left(\frac{M_{\rm stars}}{10^{11}M_{\odot}}\right) + (8.05 \pm 0.18)
\end{equation}
\noindent with a typical scatter of $(0.38 \pm 0.13)$\,dex.

If we instead estimate $M/L$ using the prescriptions of \citet{into13}, we
find the best fit to be:
\begin{equation}
\log \frac{M_{\rm BH}}{M_{\odot}} = (1.34 \pm 0.55) \log \left(\frac{M_{\rm stars}}{10^{11}M_{\odot}}\right) + (8.49 \pm 0.41)
\end{equation}
\noindent with a typical scatter of $(0.49 \pm 0.17)$\,dex.  These
relationships are displayed in Figure~\ref{fig:stmass}.

Interestingly, the \mmstars\ relationship based on the \citet{into13}
$M/L$ values is similar to that found by \citet{reines15} for
inactive black holes residing in ellipticals and classical bulges.
This would seem to contradict their finding that active galaxies lie
below the relationship defined by local quiescent galaxies, although a
direct comparison is somewhat difficult given that they used $M/L$
prescriptions of \citet{zibetti09}, who employ a different initial
mass function than \citet{into13}.

We therefore recalculated the \mmstars\ relationship for local
quiescent galaxies based on the absolute $V$ magnitudes and the
$V-K_s$ colors tabulated by \citet{kormendy13}, using both the
\citet{bell01} and \citet{into13} $M/L$ prescriptions for direct
comparison with the active galaxies in our sample.  The best fits are
shown as the dashed lines in Figure~\ref{fig:stmass}.  Using the
\citet{bell01} prescription, we find a nearly identical slope for the
quiescent galaxies compared to the active galaxies, but an intercept
that is 0.75\,dex higher, supporting the findings of \citet{reines15}
that active galaxies fall below quiescent galaxies in this parameter
space.  However, using the \citet{into13} prescription instead, we
find a slightly steeper slope for the quiescent galaxies which, when
coupled with the intercept, show the two samples to be in general
agreement at the low-mass end while diverging at the high mass end.

If we again combine the active sample with the quiescent galaxies and
the megamasers, we find a best-fit relationship of
\begin{equation}
\log \frac{M_{\rm BH}}{M_{\odot}} = (1.84 \pm 0.25) \log \left(\frac{M_{\rm stars}}{10^{11}M_{\odot}}\right) + (8.40 \pm 0.09)
\end{equation}
\noindent with a scatter of ($0.44 \pm 0.10$)\,dex. This fit is
denoted with the red long-short dashed line in the left panel of
Figure~\ref{fig:stmass}.  Once again, the consistency with the results
derived solely from the active galaxies seems to indicate that all of
these subsamples may be drawn from the same parent population.

Recently, \citet{shankar16} investigated the potential for selection
bias among the quiescent galaxy sample using Monte Carlo simulations
and a large sample of galaxies drawn from the Sloan Digital Sky
Survey.  They concluded that the quiescent galaxy sample is selected
from an upper ``ridgeline'' in the distribution of normal galaxy
properties, leading to a bias of a factor of $\sim 3$ in the
normalization of the \msigma\ relationship.  If such a bias exists,
that would argue against our choice of scaling factor for
reverberation-based masses in the active galaxy sample, and would
instead argue for $f \approx 1$.  Interestingly, when we adopt $f=1$
for the scaling of reverberation-based \mbh, and examine the
\mmstars\ relationship based on the $M/L$ prescriptions of
\citet{bell01}, we find that the active galaxy sample closely follows
the predicted unbiased relationship in Equation~6 of
\citet{shankar16}.  The stellar masses predicted by \citet{into13},
however, are undermassive compared to the predicted relationship, even
when accounting for the slight differences in assumed IMF.  However,
it appears that bars will affect the measurements of effective radii
\citep{meert15} and possibly velocity dispersion \citep{batiste17}
that are adopted for the ``unbiased'' SDSS sample considered by
\citet{shankar16}.  These effects will be strongest at the low-mass
end, where most of the active galaxies in our sample are found, thus
complicating the interpretation for reverberation-based masses.
Furthermore, forward modeling of velocity-resolved reverberation
signals by \citet{pancoast14} and \citet{grier17} has constrained the
geometry and kinematics of the broad line region and the black hole
mass for 9 AGNs, independent of any $f$ factor.  Both studies recover
modeling-based black hole masses that agree well with \mbh\ values
derived from traditional reverberation analysis and the use of
$f\approx4$ (as described in Section 4.5).  These findings argue
against the use of $f=1$ for the proper scaling of reverberation
masses, but do not rule out that there may be biases present in the
quiescent galaxy sample.

\begin{figure*}
\plotone{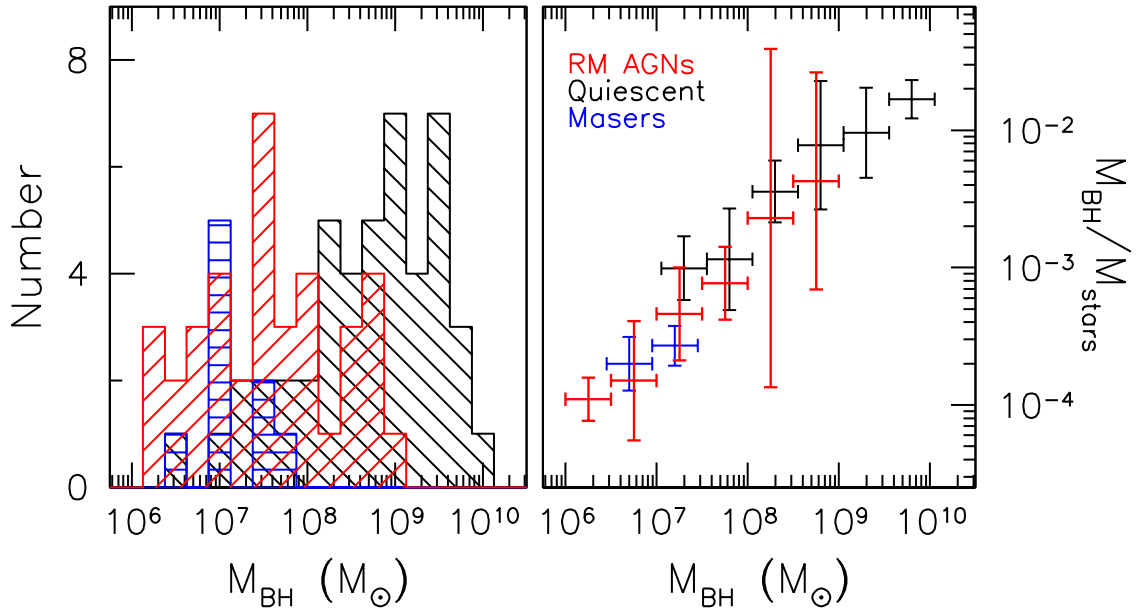}
\caption{{\it Left:} Histogram of black hole masses from the active
  sample (red), the quiescent sample (black), and the megamaser sample
  (blue). The majority of the overlap for the active and quiescent
  samples occurs at $10^7 \leq M_{\rm BH}/M_{\odot} \leq 10^9$, while
  the megamasers are completely contained within the range of black
  hole masses probed by the active sample. {\it Right:} Median black
  hole mass fraction as a function of black hole mass for the active
  sample (red), the quiescent galaxy sample (black), and the megamaser
  sample (blue).  At least three objects contribute to each of the
  bins.  The error bars on $M_{\rm BH}/M_{\rm stars}$ show the
  standard deviation for the galaxies in the bin, while the error bars
  on black hole mass show the bin size. The bins for the quiescent
  galaxy sample and the megamaser sample have been slightly offset in
  \mbh\ for clarity.} 
\label{fig:mbhbin}
\end{figure*}

Unlike for the \mmbulge\ relationship, comparisons with simulated
galaxies are much simpler when the entire galaxy stellar mass is used
because the issues with bulge-disk decompositions are avoided,
although the caveats related to the tuning of parameters in the
simulations remain.  \citet{mutlupakdil18} recently examined the
relationships between black holes and large-scale galaxy properties
for $z=0$ spiral galaxies in the Illustris simulations.  Using the
same IMF corrections described in the previous section, we compared
our best-fit \mmstars\ relationships to theirs (blue long-dashed lines
in Figure~\ref{fig:stmass}) and we find excellent agreement,
especially when we adopt the \citet{bell01} $M/L$ prescriptions, which
may argue against any potential bias in the reverberation-based
\mbh\ scaling.  \citet{volonteri16} examined the
\mmstars\ relationship for galaxies from the Horizon-AGN simulation,
and find a slope that is somewhat shallower than we have found,
although the low-mass end of their relationship may be biased by
incompleteness. \citet{steinborn15} used the Magneticum Pathfinder
Simulations to examine the \mmstars\ relationship, excluding simulated
galaxies for which $M_{\rm BH} < 5\times 10^7$\,M$_{\rm BH}$.  Perhaps
unsurprisingly, their reported best-fit relationship (with a slope of
1.09) agrees with the most massive black holes in the active galaxy
sample ($M_{\rm BH} \gtrsim 10^8$\,M$_{\odot}$), but diverges at lower
black hole masses, predicting a larger $M_{\rm BH}$ at fixed $M_{\rm
  stars}$, similar to the findings of \citet{volonteri16}.

Many large photometric surveys that are currently in operation or are
upcoming will collect photometry in multiple filters.  When
considering that these surveys that may need to be treated in an
automated fashion, the stellar mass of the galaxy based on its color
appears to be a better predictor of black hole mass than the total
galaxy luminosity in a single filter.  This can be seen from the
decreased scatter in the \mmstars\ relationship for the combined
active, quiescent, and megamaser samples ($0.44\pm0.10$\,dex) relative
to the \mlgalaxy\ relationship ($\sim0.53\pm0.09$\,dex).

\subsection{Black Hole Mass Fraction}

Finally, we also investigated the typical fraction of black hole mass
to stellar mass for each galaxy.  We find a median value of $M_{\rm
  BH}/M_{\rm stars}= 0.0005 \pm 0.0049$, however we also find a very
clear relationship between $M_{\rm BH}/M_{\rm stars}$ and \mbh, while
there appears to be no obvious relationship between $M_{\rm BH}/M_{\rm
  stars}$ and $M_{\rm stars}$ (see Figure~\ref{fig:mbhfrac}).

For comparison, we derived the black hole mass fractions for the
quiescent galaxy sample of \citet{kormendy13} and find a median value
of $M_{\rm BH}/M_{\rm stars}= 0.0058 \pm 0.0077$.  At first glance,
this would appear to demonstrate that active galaxies host
undermassive black holes compared to quiescent galaxies.  However, the
samples cover different \mbh\ ranges, with the active galaxy sample
skewed toward lower \mbh, while the quiescent galaxy sample is skewed
to higher \mbh\ (see Figure~\ref{fig:mbhbin}), and $M_{\rm BH}/M_{\rm
  stars}$ seems to depend strongly on $M_{\rm BH}$.  For a better
comparison between the two samples, we binned the galaxies in each
sample by \mbh\ with bins of width 0.5\,dex.  For each bin with three
or more objects, we computed the median black hole mass to stellar
mass fraction.  Figure~\ref{fig:mbhbin} shows the median $M_{\rm
  BH}/M_{\rm stars}$ as a function of \mbh\ for the two samples, with
the active sample in red and the quiescent sample in black.  The
majority of the overlap between the samples exists within the range
$10^7 <M_{\rm BH}/M_{\odot} < 10^9$, with the range extending to lower
black hole masses in the active galaxy sample, and extending to higher
black hole masses in the quiescent galaxy sample.  We have adopted
$M_{\rm stars}$ based on the $M/L$ predictions of \citet{bell01} in
Figure~\ref{fig:mbhbin}, but while the values slightly change, the
overall trend is the same if we adopt $M_{\rm stars}$ based on the
\citet{into13} $M/L$ values.  The two samples show broad agreement,
both in the overall trend -- with more massive black holes comprising
larger mass fractions of their galaxies -- and with the typical values
for the black hole mass fraction at a fixed value of \mbh.  While
there seems to be a tendency for the active galaxies to lie slightly
below the quiescent galaxies in the expected black hole mass fraction
at a fixed black hole mass, the values agree within the standard
deviation for each bin, and the small and uneven number of objects in
each bin make it difficult to draw firm conclusions about any apparent
offset between the two samples.  Notably, the trend appears to
continue across the full range of black hole masses probed by either
sample.

We also examined the megamaser sample of \citet{lasker16} for
comparison.  Adopting the same bins for the megamaser sample as for
the above two samples, we show the median $M_{\rm BH}/M_{\rm stars}$
in blue in Figure~\ref{fig:mbhbin}.  There is no apparent offset
between the megamaser sample and the reverberation sample, nor with
the extension of the quiescent sample to lower black hole masses.
\citet{lasker16} noted in their study that the megamaser galaxies
appeared to probe a lower $M_{\rm BH}$ at fixed galaxy mass than the
reverberation sample (as reported by \citealt{bentz09b}), but this
discrepancy has been completely erased with the larger sample and
extended range of $M_{\rm BH}$ and galaxy properties presented here.

The scaling of $M_{\rm BH}/M_{\rm stars}$ as a function of \mbh\ was
previously noticed by \citet{trakhtenbrot10}.  Using large samples of
local non-AGN galaxies and AGN galaxies at $z \approx 0.15, 1, 2$ and
scaling relationships to predict $M_{\rm stars}$ and $M_{\rm BH}$,
they found that $M_{\rm BH}/M_{\rm stars} \propto (0.7 \pm 0.1)M_{\rm
  BH}$.  A formal fit to the active, quiescent, and maser galaxies
examined here finds:
\begin{equation}
\log \frac{M_{\rm BH}}{M_{\rm stars}} = (0.71 \pm 0.04) \log \left(\frac{M_{\rm BH}}{10^{8}M_{\odot}}\right) - (2.80 \pm 0.04)
\end{equation}
\noindent with a typical scatter of $(0.04 \pm 0.02)$\,dex, which
agrees well with a formal fit to the active galaxies alone, and to the
estimated slope reported by \citet{trakhtenbrot10}.

Interestingly, we find the same scaling between $M_{\rm BH}/M_{\rm
  stars}$ and $M_{\rm BH}$ among simulated galaxies from
Illustris. \citet{vogelsberger14} provide black hole masses and galaxy
stellar masses for two subsamples of representative ``red'' and
``blue'' galaxies from the Illustris simulation. The ``blue'' galaxies
preferentially occupy the lower $M_{\rm BH}$ range that is probed
here by the active and megamaser samples, and the ``red'' galaxies
preferentially occupy the upper $M_{\rm BH}$ range probed by the
quiescent galaxies.  The scaling in $M_{\rm BH}/M_{\rm stars}$ as a
function of $M_{\rm BH}$ in the simulated galaxies matches the
observed galaxies extremely well in both slope and offset.

It is clear from these studies that the commonly-used assumption of a
constant $M_{\rm BH}/M_{\rm stars}$ is incorrect in the local universe
and possibly up to $z\approx2$. Attempts to search for cosmic
evolution of black holes and host galaxies, or to search for
differences in the evolutionary paths of distinct galaxy samples,
should be careful to account for this scaling when the samples are not
matched in $M_{\rm BH}$.

We suggest that the physical meaning of this scaling may be related to
differences in feedback efficiency as a function of galaxy mass.
High-resolution and zoom-in simulations of individual galaxies show
that supernova feedback is extremely effective at prohibiting black
hole growth at early times (e.g.,
\citealt{dubois15,trebitsch18,anglesalcazar17}).  Once the host galaxy
reaches a critical mass ($M_{\rm stars} \approx 10^9-10^{10}
M_{\odot}$; \citealt{dubois15,anglesalcazar17}), supernova feedback can
no longer restrict the gas flow to the nucleus and the black hole will
undergo a period of rapid growth, effectively ``catching up'' with the
galaxy.  This period of rapid growth is short-lived, however, because
AGN feedback soon becomes important and the black hole then regulates
its own growth and the continued growth of the galaxy (e.g.,
\citealt{dubois15,mcalpine17}).  In this scenario, we may currently be
witnessing the rapid growth phase for low-mass black holes in the
local universe.

\section{Summary}

Using high-resolution optical \hst\ images and deep, ground-based
near-infrared images, we have constrained the photometric properties
of 37 active galaxies hosting black holes with reverberation-based
\mbh\ measurements. We have compared our results with those of
megamaser galaxies and of quiescent galaxies with black hole masses
from dynamical modeling, and we have re-examined several black
hole-galaxy scaling relationships.  In general, we find that
megamasers behave as a subset of the active galaxy sample, and there
is evidence that the active and megamaser samples may be drawn from
the same parent population as the quiescent galaxies. We also find the
following:
\begin{itemize}
\item The \mlbulge\ relationship for active galaxies is slightly
  steeper in the near-infrared than the optical, and both bandpasses
  exhibit similar scatter.  There is general agreement with our
  results and those found for quiescent galaxies by \citet{kormendy13}
  and the megamaser sample of \citet{lasker16}.  $L_{\rm bulge}$ is
  found to have the tightest correlation with \mbh\ of the
  relationships examined here, and will provide the least biased
  \mbh\ estimates from photometry.
\item The \mlgalaxy\ relationship for active galaxies is only slightly
  less well defined than the \mlbulge\ relationship, but when
    combined with the megamaser and quiescent galaxy samples, the
    scatter increases significantly.  Large photometric surveys may
    forego bulge-disk decompositions and estimate unbiased black hole
    masses more quickly with total galaxy luminosity, ignoring galaxy
    morphology, but with a loss of accuracy.
\item The \mmbulge\ relationship for active galaxies is linear, while
  the quiescent galaxy sample displays a steeper slope.  Both samples
  agree at the low mass end, and the agreement is better when the
  $M/L$ prescriptions of \citet{bell01} are used rather than those of
  \citet{into13}.  The best-fit relationship for the combined active,
  megamaser, and quiescent samples agrees well with the relationship
  for the active galaxies alone, which also agrees well with the
  expectations from the high-resolution Illustris hydrodynamical
  simulations.  Agreement with other simulations is less clear because
  of incompleteness at $M_{\rm BH} \lesssim 5 \times
  10^7$\,M$_{\odot}$.
\item The active galaxy \mmstars\ relationship tends to lie slightly
  below that of the quiescent galaxy sample, but there is excellent
  agreement with the best fit for the combined active, quiescent, and
  megamaser samples.
  There is also excellent agreement between the best-fit
  \mmstars\ relationship for the active galaxies and the expectations
  from the high-resolution Illustris hydrodynamical simulations, but
  incompleteness affects comparsions with other simulations.  Large
  photometric surveys with multiple filters will achieve better
  accuracy in predicted black hole masses using the stellar mass of
  the galaxy (based on a two filter color) than the galaxy luminosity
  in a single filter.
\item The fraction of the black hole mass to the galaxy stellar mass
  is a strong function of black hole mass (but not stellar mass), with
  more massive black holes occupying larger fractions of $M_{\rm
    BH}/M_{\rm stars}$. The same trend is seen in the quiescent galaxy
  and megamaser samples, and the median black hole mass fractions at
  fixed black hole mass are similar between all three samples. The
  median value of the black hole mass fraction ranges from $\sim
  0.01$\% at $10^6$\,M$_{\odot}$ to $\sim 1.0$\% at
  $10^{10}$\,M$_{\odot}$ and follows the form $M_{\rm BH}/M_{\rm
    stars} \propto M_{\rm BH}^{0.71 \pm 0.04}$.  Studies that seek to
  compare different galaxy samples should be careful to account for
  this effect if the samples are not matched in black hole mass.
\end{itemize}

\acknowledgements

Dedicated to Marla: you were instrumental in helping a young girl to
find her voice and to not be afraid of using it. You will be sadly
missed.

We thank the anonymous referee for helpful suggestions that improved
the content and presentation of this paper.  We also thank Rachel
Kuzio de Naray, Monica Valluri, and Vardha Bennert for helpful
comments.  MCB gratefully acknowledges support from the National
Science Foundation through CAREER grant AST-1253702, and through
grants HST GO-11661 and HST GO-13816 from the Space Telescope Science
Institute, which is operated by the Association of Universities for
Research in Astronomy, Inc., under NASA contract NAS5-26555. Based
in part on observations at Kitt Peak National Observatory, National
Optical Astronomy Observatory (NOAO Prop.\ ID: 2011B-0120; PI: Bentz;
NOAO Prop.\ ID: 2013A-0438; PI: Manne-Nicholas), which is operated by
the Association of Universities for Research in Astronomy (AURA) under
a cooperative agreement with the National Science Foundation.  This
publication makes use of data products from the Two Micron All Sky
Survey, which is a joint project of the University of Massachusetts
and the Infrared Processing and Analysis Center/California Institute
of Technology, funded by the National Aeronautics and Space
Administration and the National Science Foundation.


\clearpage


\clearpage

\begin{deluxetable*}{lcccccl}
\renewcommand{\arraystretch}{1.25}
\tablecolumns{7}
\tablewidth{0pt}
\tablecaption{Galaxy Sample and Observations}
\tablehead{
\colhead{Object} &
\colhead{RA} &
\colhead{Dec} &
\colhead{$z$} &
\colhead{Date} &
\colhead{Exp Time} &
\colhead{Obs Setup}\\
\colhead{} &
\colhead{(hh:mm:ss)} &
\colhead{(\degr:\arcmin:\arcsec)} &
\colhead{} &
\colhead{(yyyy$-$mm$-$dd)} &
\colhead{(s)} &
\colhead{}
}
\startdata
Mrk\,335       &  $00:06:20.2$  &  $+20:12:10$  & 0.0258 & $2011-09-20$ & 350.0   & WIYN WHIRC H  \\  
               &                &               &        & $2006-08-24$ & 2040.0  & ACS HRC F550M \\
Mrk\,1501      &  $00:10:31.3$  &  $+10:58:30$  & 0.0893 & $2011-09-20$ & 1400.0  & WIYN WHIRC H  \\  
               &                &               &        & $2014-10-23$ & 2236.0  & WFC3 UVIS2 F547M \\
PG\,0026+129   &  $00:29:14.1$  &  $+13:16:03$  & 0.1420 & $2006-06-24$ & 2559.8  & NICMOS NIC2 F160W \\  
               &                &               &        & $2007-06-06$ & 1445.0  & WFPC2 F547M \\
Mrk\,590       &  $02:14:33.3$  &  $-00:46:00$  & 0.0264 & $2012-01-13$ & 1500.0  & WIYN WHIRC H  \\  
               &                &               &        & $2003-12-18$ & 1020.0  & ACS HRC F550M \\
3C\,120        &  $04:33:11.1$  &  $+05:21:16$  & 0.0330 & $2011-09-20$ & 200 .0  & WIYN WHIRC H  \\  
               &                &               &        & $2003-12-05$ & 1020.0  & ACS HRC F550M \\
Akn\,120       &  $05:16:11.1$  &  $-00:08:59$  & 0.0327 & $2012-01-13$ & 1000.0  & WIYN WHIRC H  \\  
               &                &               &        & $2006-10-30$ & 2040.0  & ACS HRC F550M \\
Mrk\,6         &  $06:52:12.1$  &  $+74:25:37$  & 0.0188 & $2012-01-13$ & 720.0   & WIYN WHIRC H  \\  
               &                &               &        & $2014-11-06$ & 2620.0  & WFC3 UVIS2 F547M \\
Mrk\,79        &  $07:42:33.3$  &  $+49:48:35$  & 0.0222 & $2012-01-13$ & 4140.0  & WIYN WHIRC H  \\  
               &                &               &        & $2006-11-08$ & 2040.0  & ACS HRC F550M \\
PG\,0844+349   &  $08:47:42.4$  &  $+34:45:04$  & 0.0640 & $2006-10-01$ & 2559.8  & NICMOS NIC2 F160W \\  
               &                &               &        & $2004-05-10$ & 1020.0  & ACS HRC F550M \\
Mrk\,110       &  $09:25:13.1$  &  $+52:17:11$  & 0.0353 & $2013-04-26$ & 3500.0  & WIYN WHIRC H  \\  
               &                &               &        & $2004-05-28$ & 1020.0  & ACS HRC F550M \\
NGC\,3227      &  $10:23:31.3$  &  $+19:51:54$  & 0.0039 & $2013-04-26$ & 1470.0  & WIYN WHIRC H  \\  
               &                &               &        & $2010-03-29$ & 2250.0  & WFC3 UVIS2 F547M \\
NGC\,3516      &  $11:06:47.5$  &  $+72:34:07$  & 0.0088 & $2012-01-13$ & 1625.0  & WIYN WHIRC H  \\  
               &                &               &        & $2009-11-10$ & 2660.0  & WFC3 UVIS2 F547M \\
SBS\,1116+583A &  $11:18:58.6$  &  $+58:03:24$  & 0.0279 & $2013-04-28$ & 6300.0  & WIYN WHIRC H  \\  
               &                &               &        & $2010-06-06$ & 2510.0  & WFC3 UVIS2 F547M \\
Arp\,151       &  $11:25:36.4$  &  $+54:22:57$  & 0.0211 & $2013-04-28$ & 3780.0  & WIYN WHIRC H  \\  
               &                &               &        & $2010-04-09$ & 2450.0  & WFC3 UVIS2 F547M \\
Mrk\,1310      &  $12:01:14.1$  &  $-03:40:41$  & 0.0196 & $2013-04-27$ & 4500.0  & WIYN WHIRC H  \\  
               &                &               &        & $2009-12-02$ & 2240.0  & WFC3 UVIS2 F547M \\
NGC\,4051      &  $12:03:10.1$  &  $+44:31:53$  & 0.0023 & $2013-04-26$ & 3060.0  & WIYN WHIRC H  \\  
               &                &               &        & $2010-07-17$ & 2340.0  & WFC3 UVIS2 F547M \\
NGC\,4151      &  $12:10:33.3$  &  $+39:24:21$  & 0.0033 & $2013-04-27$ & 1005.0  & WIYN WHIRC H  \\  
               &                &               &        & $2010-07-03$ & 2310.0  & WFC3 UVIS2 F547M \\
Mrk\,202       &  $12:17:55.6$  &  $+58:39:35$  & 0.0210 & $2013-04-28$ & 4800.0  & WIYN WHIRC H  \\  
               &                &               &        & $2010-04-14$ & 2510.0  & WFC3 UVIS2 F547M \\
NGC\,4253      &  $12:18:27.3$  &  $+29:48:46$  & 0.0129 & $2012-01-13$ & 1700.0  & WIYN WHIRC H  \\  
               &                &               &        & $2010-06-21$ & 2270.0  & WFC3 UVIS2 F547M \\
PG\,1226+023   &  $12:29:07.7$  &  $+02:03:09$  & 0.1583 & $2013-04-28$ & 2250.0  & WIYN WHIRC H  \\  
               &                &               &        & $2007-01-17$ & 2040.0  & ACS HRC F550M \\
PG\,1229+204   &  $12:32:04.4$  &  $+20:09:29$  & 0.0630 & $2003-11-30$ & 2559.8  & NICMOS NIC2 F160W \\  
               &                &               &        & $2006-11-20$ & 2040.0  & ACS HRC F550M \\
NGC\,4593      &  $12:39:39.4$  &  $-05:20:39$  & 0.0090 & $2013-04-27$ & 960.0   & WIYN WHIRC H  \\  
               &                &               &        & $2010-07-10$ & 2240.0  & WFC3 UVIS2 F547M \\
NGC\,4748      &  $12:52:12.1$  &  $-13:24:53$  & 0.0146 & $2013-04-27$ & 3600.0  & WIYN WHIRC H  \\  
               &                &               &        & $2010-06-28$ & 2250.0  & WFC3 UVIS2 F547M \\
PG\,1307+085   &  $13:09:47.5$  &  $+08:19:48$  & 0.1550 & $2007-01-23$ & 2559.8  & NICMOS NIC2 F160W \\  
               &                &               &        & $2007-03-21$ & 1445.0  & WFPC2 F547M \\
Mrk\,279       &  $13:53:03.3$  &  $+69:18:30$  & 0.0305 & $2013-04-27$ & 2300.0  & WIYN WHIRC H  \\  
               &                &               &        & $2003-12-07$ & 1020.0  & ACS HRC F550M \\
PG\,1411+442   &  $14:13:48.5$  &  $+44:00:14$  & 0.0896 & $2006-11-27$ & 2559.8  & NICMOS NIC2 F160W \\  
               &                &               &        & $2006-11-11$ & 2040.0  & ACS HRC F550M \\
PG\,1426+015   &  $14:29:07.7$  &  $+01:17:06$  & 0.0866 & $2007-03-03$ & 2559.8  & NICMOS NIC2 F160W \\  
               &                &               &        & $2007-03-20$ & 1445.0  & WFPC2 F547M \\
Mrk\,817       &  $14:36:22.2$  &  $+58:47:39$  & 0.0315 & $2013-04-26$ & 3520.0  & WIYN WHIRC H  \\  
               &                &               &        & $2003-12-08$ & 1020.0  & ACS HRC F550M \\
PG\,1613+658   &  $16:13:57.6$  &  $+65:43:10$  & 0.1290 & $2011-09-19$ & 1260.0  & WIYN WHIRC H  \\  
               &                &               &        & $2006-11-12$ & 2040.0  & ACS HRC F550M \\
PG\,1617+175   &  $16:20:11.1$  &  $+17:24:28$  & 0.1124 & $2006-07-09$ & 2559.8  & NICMOS NIC2 F160W \\  
               &                &               &        & $2007-03-19$ & 1445.0  & WFPC2 F547M \\
PG\,1700+518   &  $17:01:25.3$  &  $+51:49:20$  & 0.2920 & $2006-10-05$ & 2559.8  & NICMOS NIC2 F160W \\  
               &                &               &        & $2006-11-16$ & 2040.0  & ACS HRC F550M \\
3C\,390.3      &  $18:42:09.9$  &  $+79:46:17$  & 0.0561 & $2013-04-26$ & 3060.0  & WIYN WHIRC H  \\  
               &                &               &        & $2004-03-31$ & 1020.0  & ACS HRC F550M \\
Zw\,229-015    &  $19:05:26.3$  &  $+42:27:40$  & 0.0279 & $2011-09-20$ & 2000.0  & WIYN WHIRC H  \\
               &                &               &        & $2014-11-13$ & 2320.0  & WFC3 UVIS2 F547M \\  
NGC\,6814      &  $19:42:41.4$  &  $-10:19:25$  & 0.0052 & $2011-09-20$ & 1200.0  & WIYN WHIRC H  \\  
               &                &               &        & $2010-05-06$ & 2240.0  & WFC3 UVIS2 F547M \\
Mrk\,509       &  $20:44:10.1$  &  $-10:43:25$  & 0.0344 & $2011-09-19$ & 385.0   & WIYN WHIRC H  \\  
               &                &               &        & $2007-04-01$ & 1445.0  & WFPC2 F547M \\
PG\,2130+099   &  $21:32:28.3$  &  $+10:08:19$  & 0.0630 & $2011-09-19$ & 1800.0  & WIYN WHIRC H  \\  
               &                &               &        & $2003-10-21$ & 1020.0  & ACS HRC F550M \\
NGC\,7469      &  $23:03:16.2$  &  $+08:52:26$  & 0.0163 & $2011-09-19$ & 300.0   & WIYN WHIRC H  \\  
               &                &               &        & $2009-11-11$ & 2240.0  & WFC3 UVIS2 F547M 
\label{tab:whirc.obs}                  
\enddata
\end{deluxetable*}

\clearpage

\LongTables
\begin{deluxetable*}{lccccccl}
\renewcommand{\arraystretch}{1.25}
\tablecolumns{8}
\tablewidth{0pt}
\tablecaption{Galaxy Decompositions}
\tablehead{
\colhead{Object} &
\colhead{$m_{V}$} &
\colhead{$m_{H}$} &
\colhead{$r$} &
\colhead{$n$} &
\colhead{$q$} & 
\colhead{PA} &
\colhead{Note}\\
\colhead{} &
\colhead{(mag)} &
\colhead{(mag)} &
\colhead{(arcsec)} &
\colhead{} &
\colhead{} &
\colhead{($\degr$ E of N)} &
\colhead{}
}
\startdata
Mrk\,335       & $ 16.59$  & $13.26$   & $1.59 $       &  $2.9$  &  $0.85$ &  $-72.8$  &  bulge         \\
               & $ 15.93$  & $13.88$   & $2.89 $       &  $1.0$  &  $0.98$ &  ~$66.0$  &  disk          \\
Mrk\,1501      & $ 17.78$  & $16.08$   & $0.25 $       &  $1.0$  &  $0.58$ &  ~~$1.9$  &  inner disk    \\
               & $ 17.49$  & $14.22$   & $1.64 $       &  $1.1$  &  $0.81$ &  $-73.3$  &  bulge         \\
               & $ 16.10$  & $14.72$   & $14.44$       &  $1.0$  &  $0.56$ &  ~~$1.6$  &  disk          \\
PG\,0026+129   & $ 17.22$  & $15.46$   & $1.89$        &  $4.0$  &  $0.78$ &  $-83.6$  &  bulge         \\
Mrk\,590       & $16.07$   & $12.78$   & $0.90 $       &  $1.4$  &  $0.66$ &  $-35.5$  &  bulge         \\
               & $16.07$   & $12.79$   & $1.32 $       &  $0.4$  &  $0.95$ &  $-88.2$  &  barlens       \\
               & $14.28$   & $11.03$   & $5.43 $       &  $1.0$  &  $0.89$ &  ~$35.1$  &  disk          \\
3C\,120        & $ 17.10$  & $12.76$   & $1.19 $       &  $1.4$  &  $0.92$ &  $-76.6$  &  bulge         \\
               & $ 15.92$  & $12.74$   & $6.54 $       &  $1.0$  &  $0.62$ &  $-59.1$  &  disk          \\
Akn\,120       & $ 15.15$  & $11.93$   & $0.83 $       &  $3.9$  &  $0.88$ &  $-14.5$  &  bulge         \\
               & $ 14.87$  & $12.06$   & $5.10 $       &  $1.0$  &  $0.81$ &  ~$20.8$  &  disk          \\
Mrk\,6         & $15.32$   & $11.30$   & $2.89 $       &  $1.3$  &  $0.82$ &  $-51.0$  &  bulge         \\
               & $14.24$   & $12.26$   & $15.72$       &  $1.0$  &  $0.61$ &  $-49.9$  &  disk          \\
               & \nodata   & \nodata   & $5.95,4.13$   & \nodata &  $0.13$ &  $-57.2$  &  dust lane - inner \\
Mrk\,79        & $15.72$   & $12.30$   & $2.44 $       &  $3.3$  &  $0.88$ &  ~$59.1$  &  bulge         \\
               & $15.79$   & $13.58$   & $14.92$       &  $0.3$  &  $0.11$ &  ~$58.8$  &  bar           \\
               & $14.43$   & $11.84$   & $12.36$       &  $1.0$  &  $0.79$ &  ~$29.1$  &  disk          \\
PG\,0844+349   & $17.50$   & $14.34$   & $0.86$        &  $4.0$  &  $0.86$ &  $-1.9 $  &  bulge         \\
               & $16.95$   & $14.45$   & $2.91$        &  $1.0$  &  $0.78$ &  ~$47.0$  &  disk          \\
Mrk\,110       & $17.95$   & $13.74$   & $0.43 $       &  $1.6$  &  $0.93$ &  $-50.4$  &  bulge         \\
               & $16.44$   & $13.47$   & $2.83 $       &  $1.0$  &  $0.91$ &  ~$89.5$  &  disk          \\
NGC\,3227      & $13.95$   & $9.77 $   & $2.69 $       &  $2.5$  &  $0.61$ &  $-17.1$  &  bulge         \\
               & $16.72$   & $11.84$   & $0.44 $       &  $0.3$  &  $0.52$ &  ~$43.6$  &  bar           \\
               & $11.04$   & $8.03 $   & $58.39$       &  $1.0$  &  $0.42$ &  $-26.2$  &  disk          \\
NGC\,3516      & $13.38$   & $10.10$   & $2.15 $       &  $1.3$  &  $0.79$ &  ~$53.2$  &  bulge         \\
               & $15.00$   & $12.12$   & $8.13 $       &  $0.3$  &  $0.41$ &  $-17.5$  &  bar           \\
               & $13.59$   & $10.94$   & $9.11 $       &  $0.8$  &  $0.71$ &  $-12.3$  &  barlens       \\
               & $12.55$   & $11.19$   & $32.75$       &  $1.0$  &  $0.79$ &  ~$39.2$  &  disk          \\
SBS\,1116+583A & $ 20.37$  & $15.43$   & $0.33$	       &  $1.1$  &  $0.80$ &  ~$61.5$  &  bulge         \\
               & $ 20.22$  & $14.95$   & $0.96$	       &  $0.3$  &  $0.93$ &  ~$77.1$  &  barlens       \\
               & $ 19.88$  & $14.07$   & $3.42$	       &  $0.5$  &  $0.25$ &  ~$69.2$  &  bar           \\
               & $ 17.62$  & $13.26$   & $5.00$	       &  $1.0$  &  $0.87$ &  ~$67.6$  &  disk          \\
Arp\,151       & $ 15.73$  & $12.39$   & $2.37 $       &  $4.1$  &  $0.78$ &  $-25.3$  &  bulge         \\
               & $ 16.86$  & $14.13$   & $4.44 $       &  $1.0$  &  $0.28$ &  $-22.6$  &  disk          \\
Mrk\,1310      & $ 16.43$  & $13.20$   & $1.88 $       &  $3.0$  &  $0.77$ &  $-42.4$  &  bulge         \\
               & $ 17.62$  & $14.76$   & $8.00 $       &  $1.0$  &  $1.00$ &  ~~$0.0$  &  ring          \\
               & \nodata   & \nodata   & $4.76,4.90$   & \nodata &  $0.65$ &  $-42.4$  &  ring - inner  \\
               & \nodata   & \nodata   & $0.87,4.11$   & \nodata &  $0.75$ &  $-52.2$  &  ring - outer  \\
               & $ 15.28$  & $13.15$   & $5.72 $       &  $1.0$  &  $0.73$ &  $-37.6$  &  disk          \\
NGC\,4051      & $14.70$   & $12.00$   & $0.94 $       &  $0.7$  &  $0.75$ &  $-40.2$  &  bulge         \\
               & $14.00$   & $10.14$   & $6.00 $       &  $1.6$  &  $0.46$ &  $-50.8$  &  bar           \\
               & $10.11$   & $8.43 $   & $89.67$       &  $1.0$  &  $0.70$ &  $-58.5$  &  disk          \\
NGC\,4151      & $14.91$   & $13.35$   & $0.88 $       &  $2.2$  &  $0.41$ &  ~$55.9$  &  bar           \\
               & $13.82$   & $10.42$   & $2.05 $       &  $0.9$  &  $0.91$ &  $-70.5$  &  bulge         \\
               & $12.60$   & $9.40 $   & $8.50 $       &  $0.6$  &  $0.87$ &  ~$39.5$  &  barlens       \\
               & $11.09$   & $8.77 $   & $58.39$       &  $1.0$  &  $0.61$ &  $-47.2$  &  disk          \\
Mrk\,202       & $ 17.18$  & $13.56$   & $0.54 $       &  $2.7$  &  $0.80$ &  $-54.7$  &  bulge         \\
               & $ 18.21$  & $14.90$   & $4.00 $       &  $1.0$  &  $1.00$ &  ~~$0.0$  &  ring          \\
               & \nodata   & \nodata   & $2.15,1.72$   & \nodata &  $0.89$ &  ~$17.7$  &  ring - inner  \\
               & \nodata   & \nodata   & $0.70,1.86$   & \nodata &  $0.76$ &  $-67.4$  &  ring - outer  \\
               & $ 15.64$  & $13.47$   & $6.99 $       &  $1.0$  &  $0.79$ &  $-52.9$  &  disk          \\
NGC\,4253      & $17.62$   & $14.75$   & $0.19 $       &  $0.1$  &  $0.62$ &  $-85.3$  &  nucleus       \\
               & $16.46$   & $12.97$   & $1.42 $       &  $1.1$  &  $0.56$ &  $-56.8$  &  bulge         \\
               & $14.68$   & $11.77$   & $7.82 $       &  $0.5$  &  $0.30$ &  $-71.2$  &  bar           \\
               & $13.51$   & $11.88$   & $16.17$       &  $1.0$  &  $0.84$ &  ~$83.0$  &  disk          \\
PG\,1226+023   & $ 14.75$  & $13.20$   & $0.94$	       &  $3.8$  &  $0.85$ &  ~$64.8$  &  bulge         \\
PG\,1229+204   & $ 16.13$  & $13.05$   & $2.69$        &  $4.0$  &  $0.87$ &  $-18.6$  &  bulge         \\
               & $ 16.79$  & $14.56$   & $7.48$        &  $1.0$  &  $0.32$ &  ~$34.8$  &  disk          \\
NGC\,4593      & $15.17$   & $12.47$   & $2.56 $       &  $0.1$  &  $0.73$ &  $-65.8$  &  barlens       \\
               & $13.20$   & $9.85 $   & $7.65 $       &  $1.4$  &  $0.73$ &  $-84.6$  &  bulge         \\
               & $13.66$   & $10.61$   & $35.13$       &  $0.3$  &  $0.24$ &  ~$55.8$  &  bar           \\
               & $11.51$   & $9.50 $   & $74.33$       &  $1.0$  &  $0.51$ &  ~$65.9$  &  disk          \\
NGC\,4748      & $17.36$   & $13.63$   & $0.70 $       &  $0.1$  &  $0.78$ &  ~$50.4$  &  nucleus       \\
               & $14.70$   & $10.96$   & $5.82 $       &  $2.3$  &  $0.76$ &  ~$49.2$  &  bulge         \\
               & $13.99$   & $13.77$   & $16.80$       &  $1.0$  &  $0.69$ &  ~$77.4$  &  disk          \\
PG\,1307+085   & $ 16.17$  & $13.96$   & $18.13$       &  $4.0$  &  $0.78$ &  $-67.0$  &  bulge         \\
Mrk\,279       & $ 16.25$  & $12.32$   & $1.48 $       &  $1.7$  &  $0.56$ &  ~$32.7$  &  bulge         \\
               & $ 15.16$  & $12.20$   & $5.34 $       &  $1.0$  &  $0.56$ &  ~$32.9$  &  disk          \\
PG\,1411+442   & $ 16.74$  & $13.80$   & $2.01$        &  $4.0$  &  $0.68$ &  $-10.8$  &  bulge         \\
PG\,1426+015   & $ 16.68$  & $13.88$   & $6.67$        &  $4.0$  &  $0.53$ &  $-53.5$  &  bulge         \\
               & $ 16.38$  & $13.90$   & $6.41$        &  $1.0$  &  $0.39$ &  ~$52.1$  &  disk          \\
Mrk\,817       & $17.68$   & $13.24$   & $0.90 $       &  $1.0$  &  $0.82$ &  $-40.6$  &  bulge         \\
               & $17.45$   & $14.04$   & $5.57 $       &  $0.1$  &  $0.24$ &  $-72.2$  &  bar           \\
               & $14.29$   & $11.70$   & $7.93 $       &  $1.0$  &  $0.81$ &  $-66.8$  &  disk          \\
PG\,1613+658   & $ 15.92$  & $12.81$   & $4.24$	       &  $2.4$  &  $0.79$ &  $-21.7$  &  bulge         \\
PG\,1617+175   & $ 17.33$  & $15.19$   & $1.12$        &  $4.0$  &  $0.89$ &  $-67.7$  &  bulge         \\
PG\,1700+518   & $ 17.92$  & $15.24$   & $1.98$        &  $4.0$  &  $0.66$ &  ~$37.5$  &  bulge         \\
3C\,390.3      & $ 17.18$  & $13.86$   & $0.99 $       &  $1.7$  &  $0.74$ &  ~$72.0$  &  bulge         \\
               & $ 16.86$  & $13.61$   & $2.80 $       &  $1.0$  &  $0.96$ &  $-24.8$  &  disk          \\
Zw\,229-015    & $ 17.10$  & $13.92$   & $0.75$	       &  $1.1$  &  $0.72$ &  ~$48.1$  &  bulge         \\
               & $ 16.80$  & $13.49$   & $4.34$	       &  $0.3$  &  $0.54$ &  ~$38.1$  &  bar           \\
               & $ 15.65$  & $14.74$   & $13.25$       &  $1.0$  &  $0.60$ &  ~$44.8$  &  disk          \\
               & $ 17.28$  & $13.85$   & $13.42$       &  $1.0$  &  $0.56$ &  ~$42.5$  &  ring          \\
               & \nodata   & \nodata   & $19.03,14.97$ & \nodata &  $0.70$ &  ~$43.7$  &  ring - inner  \\
               & \nodata   & \nodata   & $8.26,10.66$  & \nodata &  $0.54$ &  ~$44.0$  &  ring - outer  \\
NGC\,6814      & $17.09$   & $12.79$   & $1.91 $       &  $1.8$  &  $0.47$ &  ~$21.1$  &  inner bar     \\
               & $14.95$   & $11.27$   & $1.84 $       &  $1.7$  &  $0.94$ &  ~$10.1$  &  bulge         \\
               & $14.94$   & $10.76$   & $6.17 $       &  $0.4$  &  $0.64$ &  ~$25.7$  &  bar           \\
               & $11.22$   & $8.84 $   & $44.09$       &  $1.0$  &  $0.98$ &  ~$25.0$  &  disk          \\
Mrk\,509       & $ 15.25$  & $12.41$   & $2.19 $       &  $1.0$  &  $0.67$ &  ~$82.5$  &  disk          \\
               & $ 16.98$  & $13.52$   & $4.50 $       &  $1.0$  &  $0.44$ &  $-70.3$  &  ring          \\
               & \nodata   & \nodata   & $3.45,1.05$   & \nodata &  $0.60$ &  ~$76.2$  &  ring - inner  \\
               & \nodata   & \nodata   & $2.21,9.57$   & \nodata &  $0.36$ &  ~$21.0$  &  ring - outer  \\
PG\,2130+099   & $ 17.90$  & $13.67$   & $0.54$	       &  $5.1$  &  $0.51$ &  ~$64.2$  &  bulge         \\
               & $ 16.53$  & $13.52$   & $4.34$	       &  $1.0$  &  $0.55$ &  ~$47.1$  &  disk          \\
NGC\,7469      & $15.15$   & $11.16$   & $2.94 $       &  $0.3$  &  $0.70$ &  $-61.3$  &  bulge         \\
               & $14.92$   & $11.19$   & $4.00 $       &  $1.0$  &  $0.90$ &  ~$13.0$  &  ring          \\
               & \nodata   & \nodata   & $3.09,3.46$   & \nodata &  $0.36$ &  ~$45.8$  &  ring - inner  \\
               & \nodata   & \nodata   & $1.18,1.18$   & \nodata &  $0.76$ &  ~$58.2$  &  ring - outer  \\
               & $16.82$   & $13.32$   & $0.51 $       &  $0.2$  &  $0.67$ &  ~$68.4$  &  inner disk    \\
               & $14.12$   & $11.17$   & $10.42$       &  $0.2$  &  $0.56$ &  $-58.0$  &  inner disk    \\
               & $15.65$   & $12.73$   & $33.84$       &  $1.0$  &  $0.81$ &  $-56.9$  &  disk          \\
\label{tab:fits}                                                 
\enddata                                                         
\end{deluxetable*}                                               
                                                                 
\clearpage

\begin{deluxetable*}{lccccccccc}
\renewcommand{\arraystretch}{1.25}
\tablecolumns{10}
\tablewidth{0pt}
\tablecaption{Bulge and Galaxy Magnitudes and Luminosities}
\tablehead{
\colhead{Object} &
\colhead{$D$} &
\colhead{$V_{bulge}$} &
\colhead{$V_{galaxy}$} &
\colhead{$\log L_{bulge}$ ($V$)} &
\colhead{$\log L_{galaxy}$ ($V$)} &
\colhead{$H_{bulge}$} &
\colhead{$H_{galaxy}$} &
\colhead{$\log L_{bulge}$ ($H$)} &
\colhead{$\log L_{galaxy}$ ($H$)} \\
\colhead{} &
\colhead{(Mpc)} &
\colhead{(mag)} &
\colhead{(mag)} &
\colhead{($L_{\odot}$)} &
\colhead{($L_{\odot}$)} &
\colhead{(mag)} &
\colhead{(mag)} &
\colhead{($L_{\odot}$)} &
\colhead{($L_{\odot}$)}
}
\startdata
Mrk\,335       & ~$109.5 \pm 7.1$ & 16.59   & 15.42  & $9.409  \pm 0.098$  & $9.861  \pm 0.098$  & 13.25   & 12.76  & $10.117 \pm 0.098$  & $10.312 \pm 0.098$   \\
Mrk\,1501      & ~$402.5 \pm 7.5$ & 17.22   & 15.40  & $10.325 \pm 0.082$  & $11.055 \pm 0.082$  & 14.17   & 13.53  & $10.872 \pm 0.082$  & $11.129 \pm 0.082$   \\
PG\,0026+129   & ~$653.1 \pm 7.7$ & 17.03   & 17.03  & $10.868 \pm 0.081$  & $10.868 \pm 0.081$  & 15.30   & 15.30  & $10.840 \pm 0.081$  & $10.840 \pm 0.081$   \\
Mrk\,590       & ~$112.1 \pm 7.1$ & 16.07   & 13.89  & $9.633  \pm 0.097$  & $10.494 \pm 0.097$  & 12.76   & 10.65  & $10.333 \pm 0.097$  & $11.177 \pm 0.097$   \\
3C\,120        & ~$140.9 \pm 7.1$ & 16.38   & 14.85  & $9.730  \pm 0.091$  & $10.327 \pm 0.091$  & 12.62   & 11.86  & $10.587 \pm 0.091$  & $10.891 \pm 0.091$   \\
Akn\,120       & ~$139.6 \pm 7.1$ & 14.89   & 13.95  & $10.311 \pm 0.091$  & $10.670 \pm 0.091$  & 11.87   & 11.18  & $10.879 \pm 0.091$  & $11.154 \pm 0.091$   \\
Mrk\,6         & ~~$80.6 \pm 7.1$ & 14.97   & 13.55  & $9.770  \pm 0.111$  & $10.339 \pm 0.111$  & 11.24   & 10.86  & $10.658 \pm 0.111$  & $10.808 \pm 0.111$   \\
Mrk\,79        & ~~$94.0 \pm 7.1$ & 15.64   & 13.80  & $9.657  \pm 0.103$  & $10.375 \pm 0.103$  & 12.27   & 11.14  & $10.376 \pm 0.103$  & $10.830 \pm 0.103$   \\
PG\,0844+349   & ~$279.4 \pm 7.3$ & 17.40   & 16.35  & $9.903  \pm 0.083$  & $10.330 \pm 0.083$  & 14.20   & 13.50  & $10.543 \pm 0.083$  & $10.823 \pm 0.083$   \\
Mrk\,110       & ~$150.9 \pm 7.1$ & 18.00   & 16.21  & $9.130  \pm 0.090$  & $9.823  \pm 0.090$  & 13.74   & 12.84  & $10.198 \pm 0.090$  & $10.557 \pm 0.090$   \\
NGC\,3227      & ~~$23.5 \pm 2.4$ & 13.92   & 10.93  & $9.105  \pm 0.119$  & $10.301 \pm 0.119$  & 9.76    & 7.80   & $10.184 \pm 0.119$  & $10.970 \pm 0.119$   \\
NGC\,3516      & ~~$37.1 \pm 7.0$ & 13.30   & 11.74  & $9.755  \pm 0.182$  & $10.379 \pm 0.182$  & 10.09   & 9.34   & $10.448 \pm 0.182$  & $10.747 \pm 0.182$   \\
SBS\,1116+583A & ~$118.5 \pm 7.1$ & 18.50   & 15.48  & $8.696  \pm 0.095$  & $9.907  \pm 0.095$  & 15.42   & 12.60  & $9.315  \pm 0.095$  & $10.443 \pm 0.095$   \\
Arp\,151       & ~~$89.2 \pm 7.0$ & 15.71   & 15.39  & $9.561  \pm 0.105$  & $9.691  \pm 0.105$  & 12.38   & 12.18  & $10.287 \pm 0.105$  & $10.367 \pm 0.105$   \\
Mrk\,1310      & ~~$82.7 \pm 7.0$ & 16.37   & 14.81  & $9.230  \pm 0.109$  & $9.856  \pm 0.109$  & 13.19   & 12.41  & $9.899  \pm 0.109$  & $10.211 \pm 0.109$   \\
NGC\,4051      & ~~$17.1 \pm 3.4$ & 14.70   & 10.07  & $8.515  \pm 0.190$  & $10.368 \pm 0.190$  & 11.99   & 8.18   & $9.016  \pm 0.190$  & $10.539 \pm 0.190$   \\
NGC\,4151      & ~~$16.6 \pm 3.3$ & 13.78   & 10.72  & $8.858  \pm 0.190$  & $10.083 \pm 0.190$  & 10.41   & 8.12   & $9.622  \pm 0.190$  & $10.538 \pm 0.190$   \\
Mrk\,202       & ~~$88.9 \pm 7.1$ & 17.15   & 15.30  & $8.982  \pm 0.106$  & $9.725  \pm 0.106$  & 13.55   & 12.61  & $9.818  \pm 0.106$  & $10.193 \pm 0.106$   \\
NGC\,4253      & ~~$54.4 \pm 7.0$ & 16.44   & 13.10  & $8.835  \pm 0.138$  & $10.171 \pm 0.138$  & 12.96   & 10.86  & $9.628  \pm 0.138$  & $10.470 \pm 0.138$   \\
PG\,1226+023   & ~$735.7 \pm 7.7$ & 14.81   & 14.77  & $11.882 \pm 0.081$  & $11.882 \pm 0.081$  & 13.20   & 13.20  & $11.781 \pm 0.081$  & $11.781 \pm 0.081$   \\
PG\,1229+204   & ~$274.9 \pm 7.3$ & 16.04   & 15.59  & $10.430 \pm 0.083$  & $10.619 \pm 0.083$  & 12.91   & 12.67  & $11.045 \pm 0.083$  & $11.142 \pm 0.083$   \\      
NGC\,4593      & ~~$37.3 \pm 7.5$ & 13.17   & 11.12  & $9.812  \pm 0.192$  & $10.631 \pm 0.192$  & 9.84    & 8.66   & $10.553 \pm 0.192$  & $11.024 \pm 0.192$   \\
NGC\,4748      & ~~$61.6 \pm 7.0$ & 14.58   & 13.39  & $9.688  \pm 0.127$  & $10.165 \pm 0.127$  & 10.94   & 10.77  & $10.546 \pm 0.127$  & $10.611 \pm 0.127$   \\
PG\,1307+085   & ~$718.7 \pm 7.7$ & 16.07   & 16.07  & $11.339 \pm 0.081$  & $11.339 \pm 0.081$  & 13.81   & 13.81  & $11.514 \pm 0.081$  & $11.514 \pm 0.081$   \\
Mrk\,279       & ~$129.7 \pm 7.1$ & 16.30   & 14.83  & $9.673  \pm 0.093$  & $10.246 \pm 0.093$  & 12.31   & 11.49  & $10.638 \pm 0.093$  & $10.964 \pm 0.093$   \\
PG\,1411+442   & ~$398.2 \pm 7.4$ & 16.69   & 16.71  & $10.511 \pm 0.082$  & $10.511 \pm 0.082$  & 13.67   & 13.67  & $11.062 \pm 0.082$  & $11.062 \pm 0.082$   \\
PG\,1426+015   & ~$383.9 \pm 7.4$ & 16.55   & 15.63  & $10.543 \pm 0.082$  & $10.909 \pm 0.082$  & 13.74   & 13.00  & $11.002 \pm 0.082$  & $11.299 \pm 0.082$   \\
Mrk\,817       & ~$134.2 \pm 7.1$ & 17.69   & 14.22  & $9.123  \pm 0.092$  & $10.519 \pm 0.092$  & 13.24   & 11.37  & $10.297 \pm 0.092$  & $11.044 \pm 0.092$   \\
PG\,1613+658   & ~$588.4 \pm 7.6$ & 15.91   & 15.87  & $11.220 \pm 0.081$  & $11.220 \pm 0.081$  & 12.79   & 12.79  & $11.750 \pm 0.081$  & $11.750 \pm 0.081$   \\
PG\,1617+175   & ~$507.4 \pm 7.5$ & 17.18   & 17.18  & $10.556 \pm 0.081$  & $10.556 \pm 0.081$  & 15.05   & 15.05  & $10.722 \pm 0.081$  & $10.722 \pm 0.081$   \\
PG\,1700+518   & $1463.3 \pm 8.4$ & 17.90   & 17.92  & $11.417 \pm 0.080$  & $11.417 \pm 0.080$  & 15.08   & 15.08  & $11.616 \pm 0.080$  & $11.616 \pm 0.080$   \\
3C\,390.3      & ~$243.5 \pm 7.2$ & 17.05   & 16.09  & $9.944  \pm 0.084$  & $10.313 \pm 0.084$  & 13.82   & 12.94  & $10.576 \pm 0.084$  & $10.929 \pm 0.084$   \\
Zw\,229-015    & ~$120.2 \pm 7.2$ & 16.93   & 14.82  & $9.342  \pm 0.095$  & $10.186 \pm 0.095$  & 13.89   & 12.38  & $9.940  \pm 0.095$  & $10.547 \pm 0.095$   \\
NGC\,6814      & ~~$21.8 \pm 7.0$ & 14.47   & 10.66  & $8.823  \pm 0.289$  & $10.346 \pm 0.289$  & 11.18   & 8.47   & $9.549  \pm 0.289$  & $10.635 \pm 0.289$   \\
Mrk\,509       & ~$147.0 \pm 7.1$ & \nodata & 14.87  & \nodata             & $10.344 \pm 0.090$  & \nodata & 12.05  & \nodata             & $10.849 \pm 0.090$   \\
PG\,2130+099   & ~$274.7 \pm 7.3$ & 17.77   & 16.15  & $9.739  \pm 0.083$  & $10.395 \pm 0.083$  & 13.65   & 12.82  & $10.750 \pm 0.083$  & $11.082 \pm 0.083$   \\
NGC\,7469      & ~~$68.8 \pm 7.0$ & 16.66   & 12.29  & $8.954  \pm 0.119$  & $10.701 \pm 0.119$  & 13.29   & 9.82   & $9.702  \pm 0.119$  & $11.088 \pm 0.119$   \\
\label{tab:mags}  
\enddata          
\end{deluxetable*}

\clearpage

\begin{deluxetable*}{lcccccccc}
\renewcommand{\arraystretch}{1.5}
\tablecolumns{9}
\tablewidth{0pt}
\tablecaption{Stellar and Black Hole Masses}
\tablehead{
\colhead{} &
\multicolumn{3}{c}{Bulge} &
\colhead{} &
\multicolumn{3}{c}{Galaxy} &
\colhead{} \\
\cline{2-4}
\cline{6-8}\\
\colhead{Object} &
\colhead{$V-H$} &
\colhead{$\log M_{\rm stars}$ (Bd01)} &
\colhead{$\log M_{\rm stars}$ (IP13)} &
\colhead{} &
\colhead{$V-H$} &
\colhead{$\log M_{\rm stars}$ (Bd01)} &
\colhead{$\log M_{\rm stars}$ (IP13)} &
\colhead{$\log M_{\rm BH}$} \\
\colhead{} &
\colhead{(mag)} &
\colhead{($M_{\odot}$)} &
\colhead{($M_{\odot}$)} &
\colhead{} &
\colhead{(mag)} &
\colhead{($M_{\odot}$)} &
\colhead{($M_{\odot}$)} &
\colhead{($M_{\odot}$)} 
}
\startdata
Mrk\,335        &  3.22	& $10.22 \pm 0.23$  & 	$9.99  \pm 0.30$  &&  2.58  &  $10.17 \pm 0.23$  &  $9.78  \pm 0.30$  &  $7.230^{+0.042}_{-0.044}$  \\
Mrk\,1501       &  2.82	& $10.82 \pm 0.23$  & 	$10.49 \pm 0.30$  &&  1.64  &  $10.63 \pm 0.23$  &  $10.00 \pm 0.30$  &  $8.067^{+0.119}_{-0.165}$  \\
PG\,0026+129    &  1.38	& $10.24 \pm 0.23$  & 	$9.55  \pm 0.30$  &&  1.38  &  $10.24 \pm 0.23$  &  $9.55  \pm 0.30$  &  $8.487^{+0.096}_{-0.119}$  \\
Mrk\,590        &  3.19	& $10.42 \pm 0.23$  & 	$10.19 \pm 0.30$  &&  3.16  &  $11.25 \pm 0.23$  &  $11.01 \pm 0.30$  &  $7.570^{+0.062}_{-0.074}$  \\
3C\,120         &  3.59	& $10.83 \pm 0.23$  & 	$10.70 \pm 0.30$  &&  2.86  &  $10.86 \pm 0.23$  &  $10.54 \pm 0.30$  &  $7.745^{+0.038}_{-0.040}$  \\
Akn\,120        &  2.87	& $10.85 \pm 0.23$  & 	$10.53 \pm 0.30$  &&  2.66  &  $11.04 \pm 0.23$  &  $10.68 \pm 0.30$  &  $8.068^{+0.048}_{-0.063}$  \\
Mrk\,6          &  3.67	& $10.93 \pm 0.23$  & 	$10.82 \pm 0.30$  &&  2.62  &  $10.68 \pm 0.23$  &  $10.31 \pm 0.30$  &  $8.102^{+0.037}_{-0.041}$  \\
Mrk\,79         &  3.25	& $10.49 \pm 0.23$  & 	$10.27 \pm 0.30$  &&  2.59  &  $10.69 \pm 0.23$  &  $10.31 \pm 0.30$  &  $7.612^{+0.107}_{-0.136}$  \\
PG\,0844+349    &  3.05	& $10.58 \pm 0.23$  & 	$10.31 \pm 0.30$  &&  2.68  &  $10.72 \pm 0.23$  &  $10.36 \pm 0.30$  &  $7.858^{+0.154}_{-0.230}$  \\
Mrk\,110        &  4.12	& $10.64 \pm 0.23$  & 	$10.64 \pm 0.30$  &&  3.27  &  $10.68 \pm 0.23$  &  $10.47 \pm 0.30$  &  $7.292^{+0.101}_{-0.097}$  \\
NGC\,3227       &  4.15	& $10.64 \pm 0.23$  & 	$10.65 \pm 0.30$  &&  3.12  &  $11.03 \pm 0.23$  &  $10.78 \pm 0.30$  &  $6.775^{+0.084}_{-0.112}$  \\
NGC\,3516       &  3.18	& $10.53 \pm 0.25$  & 	$10.30 \pm 0.32$  &&  2.37  &  $10.52 \pm 0.25$  &  $10.08 \pm 0.32$  &  $7.395^{+0.037}_{-0.061}$  \\
SBS\,1116+583A  &  3.00	& $9.33  \pm 0.23$  & 	$9.05  \pm 0.30$  &&  2.79  &  $10.38 \pm 0.23$  &  $10.05 \pm 0.30$  &  $6.558^{+0.081}_{-0.088}$  \\
Arp\,151        &  3.27	& $10.40 \pm 0.23$  & 	$10.19 \pm 0.30$  &&  3.14  &  $10.44 \pm 0.23$  &  $10.19 \pm 0.30$  &  $6.670^{+0.045}_{-0.054}$  \\
Mrk\,1310       &  3.12	& $9.96  \pm 0.23$  & 	$9.71  \pm 0.30$  &&  2.34  &  $9.98  \pm 0.23$  &  $9.53  \pm 0.30$  &  $6.212^{+0.071}_{-0.089}$  \\
NGC\,4051       &  2.70	& $8.92  \pm 0.25$  & 	$8.56  \pm 0.32$  &&  1.88  &  $10.13 \pm 0.25$  &  $9.56  \pm 0.32$  &  $6.130^{+0.121}_{-0.155}$  \\
NGC\,4151       &  3.36	& $9.78  \pm 0.25$  & 	$9.59  \pm 0.32$  &&  2.59  &  $10.40 \pm 0.25$  &  $10.01 \pm 0.32$  &  $7.555^{+0.051}_{-0.047}$  \\
Mrk\,202        &  3.54	& $10.04 \pm 0.23$  & 	$9.90  \pm 0.30$  &&  2.62  &  $10.07 \pm 0.23$  &  $9.69  \pm 0.30$  &  $6.133^{+0.166}_{-0.173}$  \\
NGC\,4253       &  3.43	& $9.81  \pm 0.24$  & 	$9.64  \pm 0.31$  &&  2.20  &  $10.18 \pm 0.24$  &  $9.70  \pm 0.31$  &  $6.822^{+0.050}_{-0.057}$  \\
PG\,1226+023    &  1.20	& $11.11 \pm 0.23$  & 	$10.37 \pm 0.30$  &&  1.20  &  $11.11 \pm 0.23$  &  $10.37 \pm 0.30$  &  $8.839^{+0.077}_{-0.113}$  \\
PG\,1229+204    &  2.99	& $11.06 \pm 0.23$  & 	$10.77 \pm 0.30$  &&  2.76  &  $11.07 \pm 0.23$  &  $10.73 \pm 0.30$  &  $7.758^{+0.175}_{-0.219}$  \\
NGC\,4593       &  3.30	& $10.68 \pm 0.25$  & 	$10.48 \pm 0.32$  &&  2.43  &  $10.83 \pm 0.25$  &  $10.40 \pm 0.32$  &  $6.882^{+0.084}_{-0.104}$  \\
NGC\,4748       &  3.60	& $10.79 \pm 0.24$  & 	$10.66 \pm 0.30$  &&  2.57  &  $10.46 \pm 0.24$  &  $10.07 \pm 0.30$  &  $6.407^{+0.110}_{-0.183}$  \\
PG\,1307+085    &  1.89	& $11.11 \pm 0.23$  & 	$10.55 \pm 0.30$  &&  1.89  &  $11.11 \pm 0.23$  &  $10.55 \pm 0.30$  &  $8.537^{+0.094}_{-0.161}$  \\
Mrk\,279        &  3.86	& $10.98 \pm 0.23$  & 	$10.92 \pm 0.30$  &&  3.24  &  $11.07 \pm 0.23$  &  $10.86 \pm 0.30$  &  $7.435^{+0.099}_{-0.133}$  \\
PG\,1411+442    &  2.83	& $11.01 \pm 0.23$  & 	$10.69 \pm 0.30$  &&  2.83  &  $11.01 \pm 0.23$  &  $10.69 \pm 0.30$  &  $8.539^{+0.125}_{-0.169}$  \\
PG\,1426+015    &  2.60	& $10.87 \pm 0.23$  & 	$10.48 \pm 0.30$  &&  2.42  &  $11.10 \pm 0.23$  &  $10.67 \pm 0.30$  &  $9.007^{+0.106}_{-0.164}$  \\
Mrk\,817        &  4.38	& $10.84 \pm 0.23$  & 	$10.91 \pm 0.30$  &&  2.76  &  $10.97 \pm 0.23$  &  $10.63 \pm 0.30$  &  $7.586^{+0.064}_{-0.072}$  \\
PG\,1613+658    &  2.77	& $11.68 \pm 0.23$  & 	$11.34 \pm 0.30$  &&  2.77  &  $11.68 \pm 0.23$  &  $11.34 \pm 0.30$  &  $8.339^{+0.164}_{-0.271}$  \\
PG\,1617+175    &  1.87	& $10.31 \pm 0.23$  & 	$9.74  \pm 0.30$  &&  1.87  &  $10.31 \pm 0.23$  &  $9.74  \pm 0.30$  &  $8.667^{+0.084}_{-0.128}$  \\
PG\,1700+518    &  1.95	& $11.23 \pm 0.23$  & 	$10.69 \pm 0.30$  &&  1.95  &  $11.23 \pm 0.23$  &  $10.69 \pm 0.30$  &  $8.786^{+0.091}_{-0.103}$  \\
3C\,390.3       &  3.03	& $10.60 \pm 0.23$  & 	$10.33 \pm 0.30$  &&  2.99  &  $10.94 \pm 0.23$  &  $10.66 \pm 0.30$  &  $8.638^{+0.040}_{-0.046}$  \\
Zw\,229-015     &  2.95	& $9.94  \pm 0.23$  & 	$9.64  \pm 0.30$  &&  2.35  &  $10.32 \pm 0.23$  &  $9.87  \pm 0.30$  &  $6.913^{+0.075}_{-0.119}$  \\
NGC\,6814       &  3.27	& $9.67  \pm 0.29$  & 	$9.45  \pm 0.35$  &&  2.17  &  $10.34 \pm 0.29$  &  $9.85  \pm 0.35$  &  $7.038^{+0.056}_{-0.058}$  \\
Mrk\,509        &  \nodata & \nodata        &   \nodata           &&  2.71  &  $10.76 \pm 0.23$  &  $10.40 \pm 0.30$  &  $8.049^{+0.035}_{-0.035}$  \\
PG\,2130+099    &  3.98	& $11.14 \pm 0.23$  & 	$11.11 \pm 0.30$  &&  3.17  &  $11.16 \pm 0.23$  &  $10.92 \pm 0.30$  &  $7.433^{+0.055}_{-0.063}$  \\
NGC\,7469       &  3.32	& $9.84  \pm 0.23$  & 	$9.64  \pm 0.30$  &&  2.42  &  $10.88 \pm 0.23$  &  $10.45 \pm 0.30$  &  $6.956^{+0.048}_{-0.050}$  \\

\label{tab:mass}  
\enddata          
\end{deluxetable*}

\end{document}